\documentclass[aps,prb,floatfix,twocolumn,superscriptaddress]{revtex4-2}
\usepackage{amsmath,amssymb}
\usepackage{graphicx}
\usepackage{epsfig}
\usepackage{psfrag}
\usepackage[usenames]{color}
\usepackage{xcolor}
\usepackage{hyperref}
\usepackage{soul,color}
\usepackage{physics}
\usepackage{tabu}
\usepackage{multirow}
\usepackage{bbold}
\usepackage{mathtools}
\usepackage{siunitx}

\hypersetup{colorlinks=true,citecolor={blue},linkcolor={blue},urlcolor={blue}}

\begin{document}

\title{Anomalous Higgs oscillations mediated by Berry curvature and quantum metric}

\author{Kristian~Hauser~A.~Villegas} 
\email[Corresponding author: ]{kristian.villegas@ntu.edu.sg}
\affiliation{Division of Physics and Applied Physics, Nanyang Technological University, Singapore 637371.}

\author{Bo Yang}
\email[Corresponding author: ]{yang.bo@ntu.edu.sg}
\affiliation{Division of Physics and Applied Physics, Nanyang Technological University, Singapore 637371.}
\affiliation{Institute of High Performance Computing, A*STAR, Singapore, 138632.}
\date{\today}

\begin{abstract}
Higgs spectroscopy, the study of Higgs bosons of a superconductor, is an emerging field in studying superconductivity. Here we show that the Berry curvature and the quantum metric of bands play a central role in the Higgs mode generation. They allow the detection of Higgs bosons even when the conventional contribution from the band curvature vanishes. Furthermore, we show that the Higgs mode can couple to the external electromagnetic field linearly when mediated by the Berry connection. As a result, we predict the existence of a second harmonic generation, in addition to the well-known third harmonic generation. We elucidate our theory further by considering a flat band superconductor realized by the Harper-Hubbard model and show that the geometric Higgs mode is lower bounded by the band Chern number. We demonstrate in twisted bilayer graphene the existence of the geometrically induced Higgs modes when superconductivity is realised in the nearly flat band at the magic angle.
\end{abstract}

\maketitle

The Anderson-Higgs mechanism \cite{Anderson1963} and its associated Higgs mode are two of the most far-reaching concepts in the theory of superconductivity. It inspired the solution of the mass generation of the W-Z bosons in high energy physics \cite{Englert1964,Higgs1964,Guralnik1964}, which culminated in the discovery of the Higgs boson \cite{ATLAS2012,CMS2012}, six decades after its theoretical proposal. It is well studied in superfluid $^3$He, leading to the proposed existence of heavier Nambu-partner Higgs bosons in the Standard Model \cite{Volovik2014}. Except for 2H-NbSe$_2$ superconductor where the Higgs mode was found accidentally via its coupling to the charge density wave \cite{Sooryakumar1980,Sooryakumar1981,Littlewood1981,Littlewood1982}, the observation of the Higgs mode in superconductors proved to be challenging. There are two main reasons: first, the Higgs excitation is electrically neutral in the sense that there is no conventional linear coupling to the external electromagnetic field. Second, its excitation gap is in the terahertz (THz) range and reliable THz probes are only developed recently. Because of the rapid advance in THz technology, there are recent interests to study the Higgs mode in superconductors \cite{Matsunaga2014, Cea2016, Pepin2020, Shimano2020}. This leads to an emerging field of Higgs spectroscopy where the Higgs mode is used to probe superconductor properties such as the pairing symmetry, the existence of other collective modes, and the pre-formation of Cooper pairs above the critical temperature in cuprates \cite{Chu2020}.

The magic-angle twisted bilayer graphene (TBG) was recently discovered to host superconductivity from strong correlations \cite{Cao2018}. It is a narrow band superconductor with a significantly enhanced critical temperature. In addition to its rich phase diagram \cite{Lu2019,Cao2020}, the band topology and geometry in TBG have significant and non-trivial effects as shown in the studies of the superfluid weight \cite{Hu2019,Julku2020,Xie2020,Peri2020}. The Higgs mode in such systems can be illusive because previous studies focused on the single-band with quadratic electronic dispersion. In conventional theory, the Higgs mode couples non-linearly to the electromagnetic vector potential via the band curvature \cite{Tsuji2015}. The resulting experimental signature is the third harmonic generation. However, the charge density wave is also known to generate third harmonics \cite{Cea2016} making it harder to discern the origin of such signal. In the case of TBG that has near-flat bands, this band curvature-mediated coupling is expected to be small. Naively, one would thus expect that there is negligible Higgs mode in TBG.

\begin{figure}[ht]
    \centering
    \includegraphics[trim={3.0cm 0.7cm 1cm 0.8cm},clip,width=0.45\textwidth]{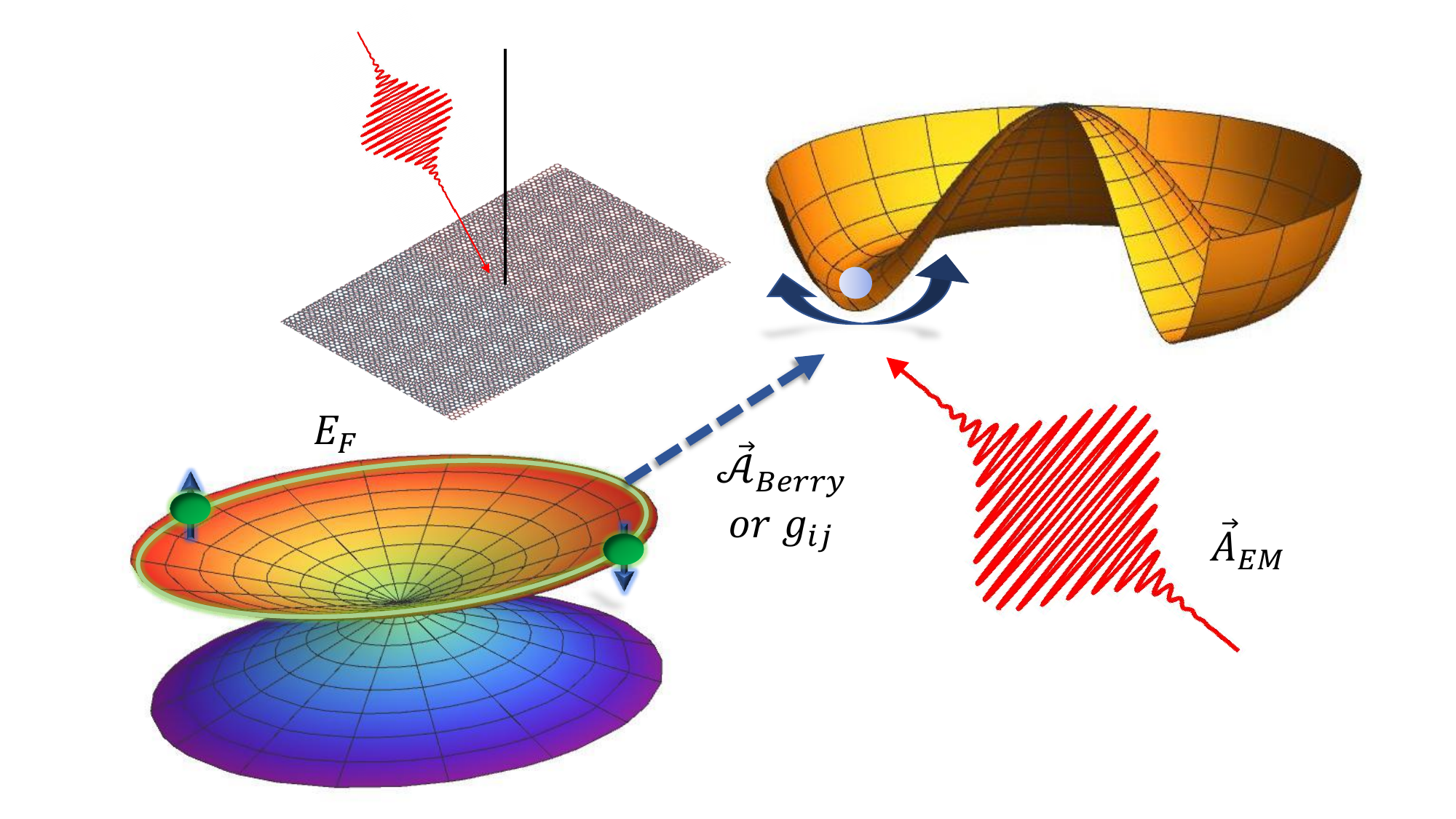}
    \caption{System schematic: An electromagnetic pulse is incident to a superconductor (upper left). The Berry curvature and the quantum metric (blue dashed arrow) mediates the excitation of Higgs mode, represented by radial oscillations about the potential minimum, by the external field (red).}
    \label{fig1}
\end{figure}

In this article we show that this is not the case. Instead, the quantum metric and the Berry curvature play significant roles in generating the Higgs mode, especially for flat band superconductors where the band curvature vanishes. The main idea is summarized in figure \ref{fig1}. We begin by formulating the general theory and deriving the multi-band pseudospin equations of motion to show the appearance of new geometric terms that are not previously accounted for. To illustrate the geometric effects more clearly, we treat the special case where the pairing occurs only in a single band. We then  elucidate our theory further by starting with two simple examples: gapped graphene and Harper-Hubbard model superconductors. The later model was studied before in the context of geometric superfluid weight \cite{Peotta2015} and Majorana-Kramers pair in cold atom systems \cite{Zeng2019}. Here we show using such model that the strength of geometric Higgs mode is bounded from below by the band Chern number. 

As an application to realistic systems accessible to experiments, we also consider the important case of TBG superconductor. We use the ten-band model, which was shown to faithfully capture the correct symmetries and fragile topology of the active bands \cite{Po2019}. There is no consensus yet as to the correct Cooper pairing mechanism for TBG. Various proposals were made \cite{Peltonen2018, Lian2019, Lewandowski2021, Cea2021, Po2018} including different pairing symmetries \cite{Fidrysiak2018, Xu2018}. We show that an external field, mediated by the Berry curvature and the quantum metric, not only significantly excites the Higgs mode in TBG, but also leads to distinct resonant signatures. Our calculation shows that the Higgs spectroscopy can be a very useful tool to experimentally probe the fundamental aspects of superconductivity in such flatband systems. Information from such experiments can in turn help us construct appropriate theoretical models capturing both the lattice and pairing symmetries in such Moire lattices.\\

\textit{General Theory--}
We start by giving a general discussion of our theory and emphasize that the framework is useful for general multiband superconductors. Consider the tight-binding Hamiltonian coupled to an external field via Peierls substitution
\begin{eqnarray}
\label{Hk}
H_K=\sum_{i\alpha,j\beta}\sum_\sigma\hat{c}^\dagger_{i\alpha\sigma}K^\sigma_{i\alpha,j\beta}e^{i\mathbf{A}\cdot(\mathbf{r}_{i\alpha}-\mathbf{r}_{j\beta})}\hat{c}_{j\beta\sigma},
\end{eqnarray}
where $i$ and $j$ label the lattice sites; $\alpha$ and $\beta$ label the orbitals; $\sigma$ denote spins; $K^\sigma_{i\alpha,j\beta}$ is the hopping amplitude; and $\mathbf{A}$ is the vector potential. The Fourier transform of this can be diagonalized: $\Tilde{K}^\sigma(\mathbf{k})=\mathcal{G}_{\mathbf{k}\sigma}\mathbb{E}_{\mathbf{k}\sigma}\mathcal{G}_{\mathbf{k}\sigma}^\dagger$, where $\mathbb{E}_{\mathbf{k}\sigma}\equiv diag(\varepsilon_{n\mathbf{k}\sigma})$ is a diagonal matrix composed of band dispersions $\varepsilon_{n\mathbf{k}\sigma}$ and $n$ labels the bands. The $n$-th column of the unitary matrix $\mathcal{G}_{\mathbf{k}\sigma}$ is the Bloch function of the $n$-th band. To account for the pairing, we use the mean field theory: 
\begin{eqnarray}
H_\Delta=-\sum_{i\alpha,j\beta}(\Delta_{i\alpha ,j\beta}\hat{c}_{i\alpha\uparrow}^\dagger\hat{c}_{j\beta\downarrow}^\dagger+H.c.)
\end{eqnarray}
For our purposes, we only consider intra-orbital pairing so that we can write the self-consistency condition as $\Delta_{i\alpha}=U\langle \hat{c}_{i\alpha\downarrow}\hat{c}_{i\alpha\uparrow}\rangle$, where $U$ is the strength of the effective electron-electron interaction. We further assume that the pairing potential has lattice-translation symmetry so that it is also diagonal in momentum space. 

The Bogoliubov-de Gennes (BdG) Hamiltonian now reads $H=\sum_\mathbf{k}\hat{\psi}^\dagger_\mathbf{k}H_\mathbf{k}(\mathbf{A})\hat{\psi}_\mathbf{k}$, 
where the Bloch Hamiltonian, upon introducing a chemical potential $\mu$, is given by
\begin{eqnarray}
\label{BdGBloch}
H_\mathbf{k}(\mathbf{A})=
\begin{pmatrix}
\mathbb{E}_{\mathbf{k}-\mathbf{A}}-\mu & \mathcal{G}^\dagger_{\mathbf{k}-\mathbf{A}}\Delta\mathcal{G}_{\mathbf{k}+\mathbf{A}}\\
\mathcal{G}^\dagger_{\mathbf{k}+\mathbf{A}}\Delta\mathcal{G}_{\mathbf{k}-\mathbf{A}} & -(\mathbb{E}_{\mathbf{k}+\mathbf{A}}-\mu)
\end{pmatrix}.
\end{eqnarray}
Here, $\Delta$ is the pairing potential matrix in the orbital space. The Nambu spinor is given by $\hat{\psi}_\mathbf{k}=(\hat{d}_{1,\mathbf{k}\uparrow},\cdot\cdot\cdot,\hat{d}_{N,\mathbf{k}\uparrow},\hat{d}^\dagger_{1,-\mathbf{k}\downarrow},\cdot\cdot\cdot,\hat{d}^\dagger_{N,-\mathbf{k}\downarrow})^T$ where $1,2,\cdot\cdot\cdot,N$ labels the bands. We focused only on the case where the particles have spin up and the holes have spin down. Expanding the diagonal block of \eqref{BdGBloch} in powers of $\mathbf{A}$ gives the conventional contribution in the Higgs generation \cite{Tsuji2015}, which has the form $\propto\frac{1}{2}(\partial_{k_i}\partial_{k_j}\varepsilon_{\mathbf{k}\alpha})A^iA^j$. This term vanishes for flat bands.

The geometric contribution to the Higgs mode comes from the pairing terms involving $\tilde{\Delta}_\mathbf{k}(\mathbf{A})=\mathcal{G}^\dagger_{\mathbf{k}-\mathbf{A}}\Delta\mathcal{G}_{\mathbf{k}+\mathbf{A}}$, which are the off-diagonal blocks of \eqref{BdGBloch}. When expanded in terms of $\mathbf{A}$, this gives the geometric terms
\begin{align}
\label{geomterms}
\Delta_0\mathcal{A}_{\mathbf{k}i\alpha}A^i,\;\; g_{\mathbf{k},ij}A^iA^j,\;\; \mathcal{A}_{\mathbf{k}i}\mathcal{A}_{\mathbf{k}j} A^iA^j,
\end{align}
where $\mathcal{A}_\mathbf{k}\equiv i\mathcal{G}^\dagger_\mathbf{k}\nabla_\mathbf{k}\mathcal{G}_\mathbf{k}$ is the Berry connection and $g_{\mathbf{k},ij}\equiv \partial_\mathbf{k_i}\mathcal{G}^\dagger_\mathbf{k}\partial_\mathbf{k_j}\mathcal{G}_\mathbf{k}-\mathcal{A}_{\mathbf{k}i}\mathcal{A}_{\mathbf{k}j}$ is the quantum metric. We introduce the generalized version of pseudospin for an $N$-band superconductor: $\mathbf{\Lambda}_\mathbf{k}\equiv\frac{1}{2}\langle\hat{\psi}^\dagger_\mathbf{k}\mathbf{\Gamma}\hat{\psi}_\mathbf{k}\rangle$, where $\{\mathbf{\Gamma}\}$ are composed of the generators of SU($2N$) and the identity matrix, and $2N$ comes from the particle and hole copies of each band. The expectation value is taken with respect to the superconducting ground state. The BdG Hamiltonian can now be written in the form 
\begin{eqnarray}
\label{hamiltonian}
H(\mathbf{A})=2\sum_\mathbf{k}\mathbf{B}_\mathbf{k}(\mathbf{A})\cdot\mathbf{\Lambda}_\mathbf{k}
\end{eqnarray}
where the pseudomagnetic field is given by $B_a(\mathbf{k},\mathbf{A})=\frac{1}{4}\Tr{\Gamma_aH_\mathbf{k}(\mathbf{A})}$. The pseudomagnetic field contains the geometric terms \eqref{geomterms} when expanded in powers of $\mathbf{A}(t)$.

The equation of motion for the pseudospin is obtained from the Heisenberg equation $\partial_t\mathbf{\Lambda}_\mathbf{k}=i[H,\mathbf{\Lambda}_\mathbf{k}]$, taking the compact form
\begin{eqnarray}
\label{eom}
\partial_t\Lambda_{\mathbf{k}a}=8f_{abc}B^b_{\mathbf{k}}(\mathbf{A})\Lambda_\mathbf{k}^c,
\end{eqnarray}
where $\{f_{abc}\}$ are the structure constants of SU($2N$). For the single band case, this reduces to the usual $\partial_t\mathbf{\sigma}_\mathbf{k}=2\mathbf{B}_\mathbf{k}\times\sigma_\mathbf{k}$ \cite{Anderson1958,Barankov2004,Yuzbashyan2005}. The equation of motion shows that the pseudomagnetic field determines the time evolution of the pseudospins. More importantly, it contains new geometric terms \eqref{geomterms} that drives, via the external vector potential, the pseudospin fluctuations.

The pseudospin fluctuations, driven by both the conventional and the new geometric terms, give rise to the Higgs mode, which can be obtained from the self-consistency condition
\begin{eqnarray}
\label{scc}
\delta\Delta(t)=U\sum_{\mathbf{k},\alpha}(\Lambda^{1\alpha}_\mathbf{k}+i\Lambda^{2\alpha}_\mathbf{k}),
\end{eqnarray}
where the superscripts in $\Lambda$s come from our splitting of the SU(2N) generators into tensor products of the Pauli matrices (particle-hole space) and the SU(N) generators (band space). For multi-band systems, the electron-electron interaction $U$ should be, in general, a matrix. For simplicity we consider only pairings in a single band, which is sufficient in illustrating the interesting geometric effects. This is justified as a typical band gap and band width is of the order eV while typical pairing potential is of the order meV.


The main message here is that when the pseudo-magnetic field in \eqref{eom} is expanded in powers of $\mathbf{A}$, it contains terms involving the band curvature $\partial_{k_i}\partial_{k_j}\varepsilon_{\mathbf{k}_F}A^iA^j$ and the geometric contributions \eqref{geomterms} involving the Berry connection and the quantum metric. The former is the conventional contribution to the Higgs mode, which vanishes for flat bands. The later geometric contributions are the main result of this work.\\ 

\textit{Example A: Gapped graphene --} We now elucidate the role of band quantum geometry to the Higgs mode generation with a simple toy model of gapped graphene s-wave superconductor. We assume that $\mu>\Delta_0$ so that the pairing only occurs in the conduction band, and consider an incident external field with frequency $\Omega$. The system of differential equations can now be solved order by order in perturbation $\mathbf{A}$ using the Laplace transformation \cite{SM}. Solving Eq.(\ref{eom}) and Eq.(\ref{scc}) gives the first order Higgs mode:
\begin{eqnarray}
\label{delta1}
\delta\Delta^{(1)}_\mathbf{p}(t)&=&\mathbb{B}^{(1)}_{geom}(t,\mathcal{A})_{ij}p^iA^j
\end{eqnarray}
where the $\mathbb{B}^{(1)}_{geom}(t,\mathcal{A})_{ij}$ matrix is a functional of the Berry connection. The appearance of the Berry connection here comes from gauge fixing where we choose the zeroth order parameter $\Delta_0$ to be real. The Higgs mode is gauge invariant as we show explicitly in the Supplementary Material (SM) \cite{SM}. 

\begin{figure}[t]
    \centering
    \includegraphics[trim={7cm 5cm 2.8cm 4cm},clip,width=0.6\textwidth]{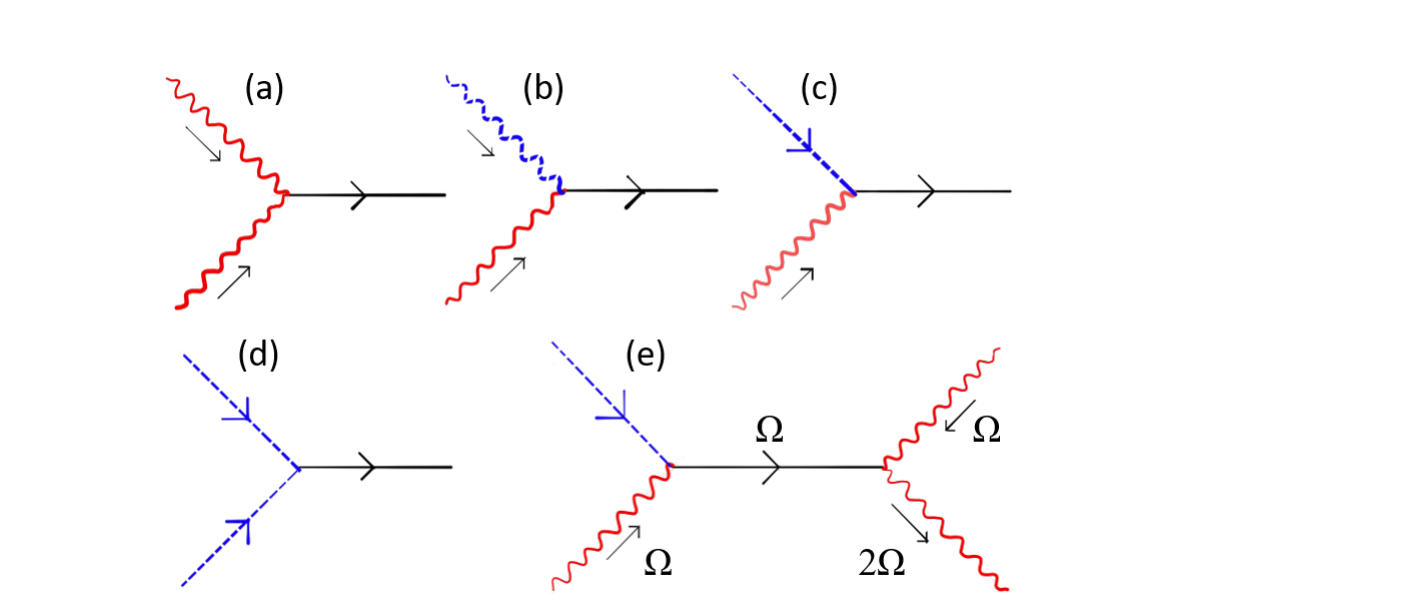}
    \caption{Feynman diagrams (a-d) for Higgs mode (black lines) generation. Red and blue dashed squiggly lines denote electromagnetic and Berry vector potentials, respectively. The straight blue dashed lines denote first order excitations. (e) Second harmonic generation.}
    \label{fig2}
\end{figure}

Notice that \eqref{delta1} is linear in $\mathbf{A}$. The coupling has a similar structure as the nonlinear case, figure \ref{fig2} (a), except that one of the vector potentials is replaced by the Berry connection as illustrated in figure \ref{fig2} (b). An important consequence of this is the generation of the second harmonics as shown in figure \ref{fig2} (e). In the conventional theory, only the third harmonic generation is possible \cite{Tsuji2015}. The second harmonic generation therefore is a unique experimental signature that we predict in our theory. The first order Higgs mode vanishes when $\mathbf{p}=0$. Hence, to observe the predicted second harmonic generation, there must be a non-zero center-of-mass momentum of the Cooper pairs. This can be achieved by inducing a supercurrent in the superconducting sample. The matrix $\mathbb{B}^{(1)}_{geom}(t,\mathcal{A})_{ij}$ have poles at $\Omega=\pm 2\Delta_0$ that are consistent with the known Higgs mode gap \cite{Anderson1958, Tsuji2015}, with the plus-minus frequency pairs coming from the particle-hole symmetry. For the rest of the examples that we consider below, we will focus on the positive frequency.

The resulting expression for the second order Higgs mode is long but can be schematically divided into two major contributions:
\begin{eqnarray}
\delta\Delta^{(2)}_\mathbf{p}(t)=[\mathbb{C}_{band}^{(2)}(t)_{ij}+\mathbb{B}_{geom}^{(2)}(t,\mathcal{A},g)_{ij}]A^iA^j.
\end{eqnarray}
The explicit forms of $\mathbb{C}_{band}^{(2)}(t)_{ij}$ and $\mathbb{B}_{geom}^{(2)}(t,\mathcal{A},g)_{ij}$ are given in \cite{[{See Supplemental Material at [URL]}]SM}. The first quantity $\mathbb{C}_{band}(t)_{ij}$ is the well-known conventional coupling \cite{Tsuji2015, Shimano2020}, which depends on the band curvature $\partial_{k_i}\partial_{k_j}\varepsilon_{\mathbf{k}_F}$.
\begin{figure*}[ht]
    \centering
    \includegraphics[trim={0.0cm 6.6cm 0.0cm 5.1cm},clip,width=1\textwidth]{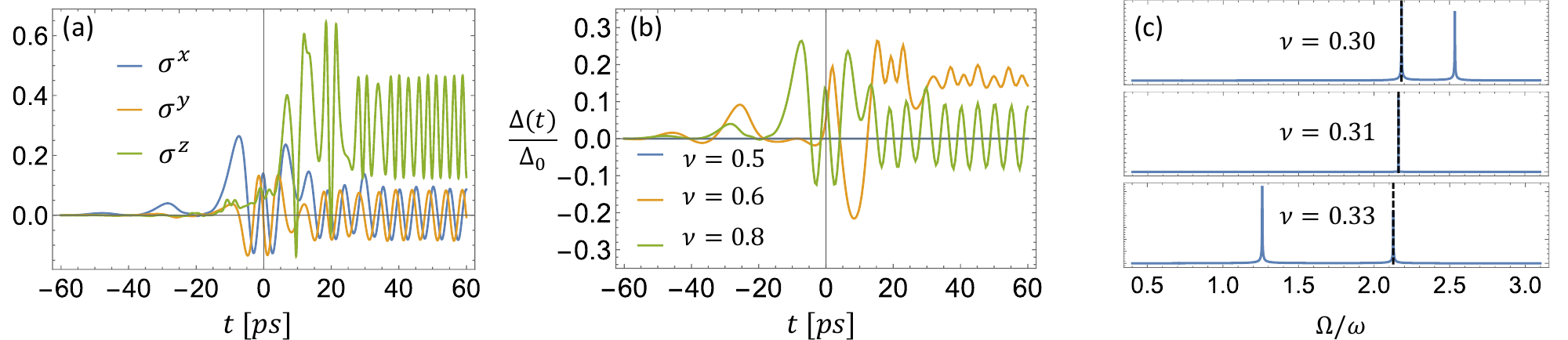}
    \caption{Pseudospin oscillations in Harper-Hubbard model (a). Higgs mode for various filling fractions $\nu$ (b). Higgs mode resonance in frequency space in units of $Un_\phi$ (c). The vertical dashed lines locate the corresponding quasiparticle excitation gap $\omega$ at the given filling fraction.}
    \label{fig3}
\end{figure*}
The term $\mathbb{B}_{geom}(t,\mathcal{A},g)_{ij}$ is a new result, which depends on the geometric quantities $\mathcal A$ and $g$. There are three resonance frequencies: $\Delta_0$, $2\Delta_0$, and  $4\Delta_0$. The first is typical of second-order interaction $\mathbf{A}(t)^2$ \cite{Tsuji2015} as shown figure \ref{fig2} (a), while the second and third resonance frequencies come from the first order excitations acting as sources for the second order correction as shown in figures \ref{fig2} (c) and (d). In contrasts to the conventional contribution, this is non-zero even in the flat band limit \cite{[{See Supplemental Material at [URL]}]SM}.\\

\textit{Example B: Harper-Hubbard model --} It is also useful to consider a model capturing the physics of strong correlation, flat multi-bands, and non-trivial band topology, but is sufficiently simple to allow analytical calculations. These are captured by the Harper-Hubbard model. While our main motivation for using this model is purely theoretical, the rapid progress of various platforms such as cold atom and Moire systems to engineer strong correlation and non-trivial band topology opens up the possibility that such model can be experimentally realized in the future. Using this model we will also show that the strength of the geometric Higgs mode is bounded from below by the band Chern number.

The Harper-Hubbard model has the hopping matrix: 
\begin{eqnarray}
K^\sigma_{ij}=-J(\rho^{-\sigma i_y}\delta_{i+\hat{x},j}+\rho^{\sigma i_y}\delta_{i-\hat{x},j}+\delta_{i+\hat{y},j}+\delta_{i-\hat{y},j})\qquad
\end{eqnarray}
where we have $\rho=\exp\{ 2\pi in_\phi\}$ with $n_\phi=1/Q$, $Q\in\mathbb{Z}^+$ the flux per plaquette and $\sigma=\pm 1$ for spin up and down, respectively. It has a self-consistent superconducting solution \cite{Peotta2015} $\varepsilon =E_{\bar{n}}-\mu=Un_\phi \left(\frac{1}{2}-\nu\right)$,
$\Delta_0 =Un_\phi\sqrt{\nu (1-\nu)}$, and
$\omega =2\sqrt{\varepsilon^2+\Delta_0^2}=Un_\phi
$.
Here $\nu$ is the band filling fraction and $U$ is the on-site Hubbard interaction. We omit the band label $\bar{n}$ in the energy $\varepsilon$ measured with respect to the chemical potential $\mu$, as we will be concerned only with one partially-filled band.

We consider the situation where the band gaps are much larger than the interaction strength $U$, which in turn is much larger than the band width. This can be satisfied for $n_\phi\ll 1$. The band is then approximately flat. We label the band where the pairing occurs by $\bar{n}$ and the $\mathbf{k}$-independent band energy by $E_{\bar{n}}$. The resulting equations of motion can be solved numerically for an external pulse field $\mathbf{A}(t)=\mathbf{A}\sin (\Omega t)e^{-(t/\tau)^2}$. The evolution of the pseudospin components is shown in figure \ref{fig3} (a) for $\tau=50$ ps and $\nu=0.2$. The Higgs oscillations are shown in figure \ref{fig3} (b) for various band filling fractions. Note that there is no Higgs oscillation at half-filling $\nu=0.5$. This comes from the identity obeyed by the pseudospins $\Delta_0\sigma^x_\mathbf{k}(t)=\epsilon\sigma^z_\mathbf{k}(t)$ and the fact that $\epsilon=0$ at $\nu=0.5$. Only $\nu\geq 0.5$ is shown as the system have a particle-hole symmetry giving identical results for $\nu\leq 0.5$. It is clear from these figures that the external field was able to drive the pseudospin and Higgs oscillations in the flatband limit where the conventional contribution vanishes.

Analytic results can be obtained for purely sinusoidal external field. In compact form:
\begin{eqnarray}
\label{deltatilde}
\tilde{\Delta}^{(2)}_R(s)&=&\frac{\alpha_1(s)G_{ij}A^iA^j+\alpha_2(s)B_{ij}A^iA^j}{(s^2+\Omega_1^2)(s^2+\Omega_2^2)(s^2+\Omega_3^2)(s^2+\Omega_4^2)},
\end{eqnarray}
where $\alpha_1(s)$ and $\alpha_2(s)$ also depend on the parameters $U$, $\nu$, and $n_\phi$. Their explicit forms are given in the SM \cite{SM}. The argument $s$ comes from the Laplace transformation $t\rightarrow s$. The tensor in the second term of the numerator is $B_{ij}=\mathcal{A}_i\mathcal{A}_j$. The interesting physics, however, comes from the quantum metric $G_{ij}$ and the denominator, which we now discuss.

The poles in \eqref{deltatilde} contain the information about the collective modes of the superconductor. They are given by: $\Omega_1=Un_\phi/3$, $\Omega_2=U\sqrt{1+(1-2\nu)^2}$, and $\Omega_3=Un_\phi$. The fourth pole $\Omega_4=2\Omega$ is due to the driving field. The resonance occurs when this driving frequency coincides with one of the collective modes. Note that since $\epsilon$ is a non-zero constant, except for half filling, the quasiparticle gap is $\omega$ and not just $2\Delta_0$. Figure \ref{fig3} (c) shows the peaks in the frequency domain, which are due to $\Omega_2$ and $\Omega_3$. The pole $\Omega_1=Un_\phi/3$ does not contribute to the peak as the numerator in \eqref{deltatilde} vanishes at this frequency. As shown in the figure, as $\nu$ is increased from $0.30$, the resonance peak due to $\Omega_2$ moves to the left and becomes subgap for $\nu=0.33$. This behavior does not occur in conventional single-band superconductors and is peculiar to the flatband Harper-Hubbard model.

The geometric contribution to the Higgs mode for the Harper-Hubbard model is proportional to the quantum metric $G_{ij}$, which in matrix form is given by $\mathbb{G}=2\pi$ diag$(2\bar{n}+1,2\bar{n}+1)$, where $\bar{n}=0,1,2,\cdot\cdot\cdot$ labels the Landau levels \cite{Peotta2015}. Hence, the coupling of the external field with the quantum metric is stronger for higher Landau levels. Using the positive semidefiniteness property of the quantum geometric tensor, this coupling with the quantum metric is bounded from below by the Chern number $C_{\bar{n}}$ of the $\bar{n}$th band: $|2\bar{n}+1|\geq C_{\bar{n}}$ \cite{SM}. An analogous bound was found previously in the context of superfluid weights  \cite{Peotta2015}. Thus the geometric strength for the Higgs mode is also lower bounded by the Chern number of the band, and is generally not small.\\

\textit{Twisted bilayer graphene --} An important application of our theory would be for the twisted bilayer graphene, which hosts superconductivity in its nearly-flat bands \cite{Cao2018}. The interactions responsible for the superconductivity must eventually come from real-space Hamiltonian \cite{Fidrysiak2018} which necessitates a minimal tight-binding model. This is furnished by the ten-band model, which faithfully captures the correct symmetries and topology of the active bands \cite{Po2019}. In this model, the $p_\pm$ and $p_z$ orbitals are attached to the triangular lattice; $s$ to the Kagome; and another $p_\pm$ to the honeycomb lattice. The band structure is shown in figure \ref{fig4} (a). 

\begin{figure*}[ht]
    \centering
  \includegraphics[trim={0.0cm 5.8cm 0.0cm 5.1cm},clip,width=1\textwidth]{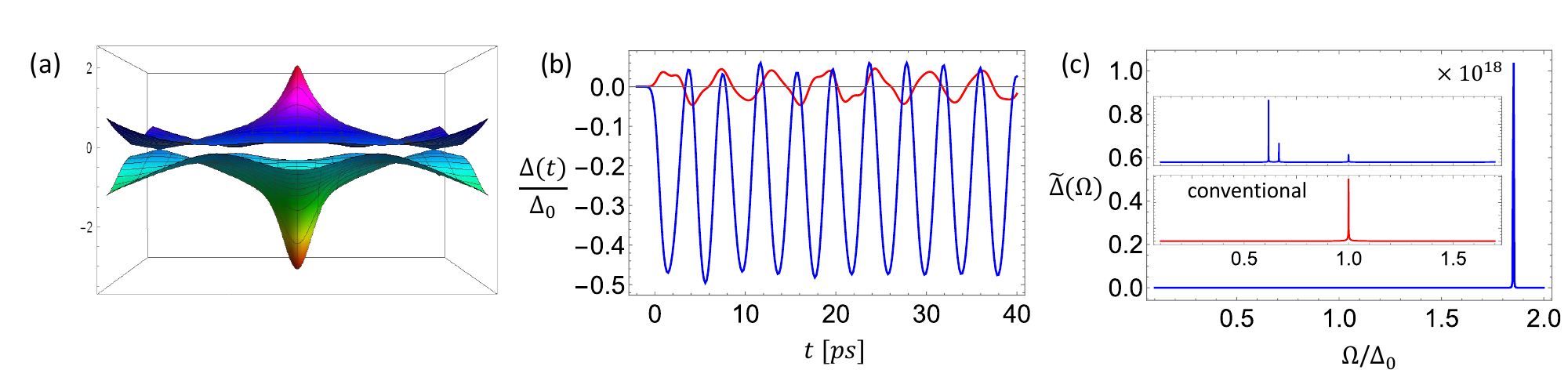}
    \caption{(a) The two active bands of TBG using the ten-band tight-binding model. (b) Geometric (blue) and conventional (red) Higgs oscillations in TBG induced by a short electromagnetic pulse of width $\tau=\pi\hbar/\Delta_0\sim 1.0$ ps and centered at $t=0$. (c) Higgs mode in frequency domain. Inset shows the conventional (lower panel) and geometric (upper panel) contributions.}
    \label{fig4}
\end{figure*}

We first consider a mean-field s-wave superconductivity \cite{Peltonen2018, Ray2019} and solve numerically the Higgs mode using a short external pulse of the form $\mathbf{A}(t)=\mathbf{A}\exp \{-\left(t/\tau\right)^2\}$, with pulse temporal width $\tau=\pi\hbar/\Delta_0$, which is typical in Higgs mode studies \cite{Chou2017}. The electric field is chosen to be linearly polarized along the x direction. We found similar results for polarizations along $\hat{x}+\hat{y}$ and $\hat{y}$ directions. We first consider the case where there is no supercurrent so that the linear contributions vanish when the contributions from the two valleys are combined. The main geometric contribution then comes from the second order $g_{ij}(\mathbf{k})A^iA^j$ and $\mathcal{A}_{\mathbf{k}i}\mathcal{A}_{\mathbf{k}j}A^iA^j$. The twisting of the Anderson pseudospins is strongest along the Fermi surface \cite{Chou2017}. Hence, in calculating the Higgs mode, we summed over $\mathbf{k}$ along the Fermi surface.

Figure \ref{fig4} (b) shows both the geometric and conventional contributions to the Higgs mode for the chemical potential $\mu=-2.0$ meV, which is just below half-filling of the lower band where the superconductivity is observed in TBG. We see that the geometric contribution is larger than the conventional one.

To gain further insight, we investigate the Higgs mode in the frequency domain. This allows us to identify the collective modes from the resonance frequencies. The equations of motion can be solved analytically for a purely sinusoidal external field $\mathbf{A}(t)=\mathbf{A}\sin (\Omega t)$. The calculation is further simplified if we assume that the pairing occurs only in the lower active band. This is justified since, while this active band is narrow $\sim 10$ meV, the pairing gap is $\lesssim 0.1$ meV and is therefore at least two orders of magnitude smaller than the bandwidth. To calculate the self-consistent equation for the Higgs mode, we only integrate within a narrow strip $\pm\omega_D$ around the Fermi surface since the pseudospins are significantly twisted only around this strip \cite{Chou2017}. If the pairing in TBG is caused by phonons \cite{Lewandowski2021, Bernevig2021}, this would be the Debye cut-off. 


The Berry connection allows the linear coupling of the pseudospins with the external vector potential. However, from the time-reversal symmetry, the Berry connection obeys $\vec{\mathcal{A}}_{+,\mathbf{k}}=-\vec{\mathcal{A}}_{-,-\mathbf{k}}$, where the subscript $\pm$ labels the valleys. Because of this, the first order Higgs contribution vanishes when summed over the two valleys. We get a non-vanishing contribution if the integration over the Fermi surface is shifted by a finite amount, say $\mathbf{p}$, for both valleys. This can be done by inducing a supercurrent in the sample so that the Cooper pairs have center-of-mass momentum $2\mathbf{p}$. 
The first order Higgs mode is now given by
\begin{align}
\Delta^{(1)}_R(s)=-\frac{8\lambda\omega_D\Omega s p^iD_{ij}A^j}{(s^2+\Omega^2)(s^2+4\Delta_0^2)}
\end{align}
where $D_{ij}=\sum_\mathbf{k}(\partial_{k_i}\varepsilon_\mathbf{k})\mathcal{A}_{\mathbf{k}j}$ with the summation over the neighborhood of the Fermi surface. In frequency domain, the pole at $\nu=\Omega$ gives the driving frequency of the external field, while the pole $\nu=2\Delta_0$ gives the gap of the Higgs excitation.

We plot the second order Higgs mode in the frequency domain in figure \ref{fig4} (c). As can be seen in the figure, there is large peak around $\sim 1.8\Delta_0$. This comes from the coupling $\Delta_I^{(1)}(t)\sigma^{z(1)}(t)$, which in turn is proportional to $(\mathcal{A}\cdot\mathbf{A})^2$ and is therefore a band-geometric contribution. Note here that we only fixed the gauge once so that the zeroth-order order parameter is real but the higher order corrections can have imaginary part. This is in contrasts to readjusting the gauge order by order to keep the order parameter real \cite{Arseev2006}. The large peak at $\sim 1.8\Delta_0$ comes from the merging of three different poles effectively giving a single third order pole. The lower panel of the inset shows the purely conventional contribution, which is the well-known Anderson pseudospin resonance, while the upper panel shows the additional peaks due to the geometric contributions. The conventional contribution at $\Omega=\Delta_0$ is relatively weak, compared to geometric contributions, as expected for a narrow-band superconductor.

We have also calculated the case of $d+id$ pairing \cite{Fidrysiak2018, Xu2018} and found distinct dominant peaks due to the geometric contributions. This case is richer than the s wave as various subgap oscillation modes exist due to symmetry quenching \cite{Schwarz2020, Barlas2013}. We leave the detailed study of this for future work.

\textit{Conclusion -- } We have shown that there are couplings involving the Berry connection, quantum metric, and external field to generate an anomalous Higgs mode beyond those that are predicted in the conventional theory. More importantly, such geometric Higgs mode exists even in the flat band limit. We demonstrated that there can be a linear coupling with the external field and, as consequence, we predict the generation of second harmonics. Using the Harper-Hubbard model, we have shown that the coupling is bounded from below by the band Chern number. We have calculated the Higgs mode for TBG and showed that the geometric contribution gives rise to a resonance distinct from the conventional one.

\begin{acknowledgments}
We would like to thank D.~M. Ca\~{n}eso, R.~C. Bernardo, and J. Song ~C.~W. for useful discussions. This work is supported by the Singapore National Research Foundation (NRF) under NRF fellowship Award NRF-NRFF12-2020-0005, a Nanyang Technological University start-up grant (NTU-SUG), and Singapore Ministry of Education MOE2018-T3-1-002.
\end{acknowledgments}
\bibliography{main}
\bibliographystyle{apsrev4-2}

\begin{widetext}

\appendix

\newpage

\section{SUPPLEMENTAL MATERIAL: Anomalous Higgs oscillations mediated by Berry curvature and quantum metric}

\maketitle

In this Supplemental Material, we show the details of our calculations and expound some discussions that were left out in the main text. Section \ref{generaltheory} gives details of our general theory. Sections \ref{gappedgraphene} and \ref{harperhubbard} give the details of our calculation for the gapped graphene and Harper-Hubbard model examples, respectively. Our calculations for TBG is given in Section \ref{tbg}. Lastly, Section \ref{gaugeinvariance} gives a detailed discussion of the gauge invariance. There we show that the gauge invariance of the Higgs mode is satisfied order by order in the expansion of the external field. 

\section{General Theory.} 
\label{generaltheory}
We assume that the electromagnetic perturbation is sufficiently weak. The expansion of the matrix whose column elements are composed of the Bloch functions is given by
\begin{eqnarray}
\label{Gexpand}
\mathcal{G}^{(\dagger)}_{\mathbf{k}\pm\mathbf{A}}=\mathcal{G}^{(\dagger)}_\mathbf{k}\pm\partial_i\mathcal{G}^{(\dagger)}_\mathbf{k}A^i+\frac{1}{2}\partial_i\partial_j\mathcal{G}^{(\dagger)}_\mathbf{k}A^iA^j+\cdot\cdot
\end{eqnarray}
where the partial derivatives are understood to be taken with respect to $\mathbf{k}$.

For the diagonal block of the BdG Hamiltonian, that is, the kinetic part, we have
\begin{eqnarray}
\mathbb{E}_{\mathbf{k}\pm\mathbf{A}}\approx[\varepsilon_{\mathbf{k}\alpha}\pm\partial_j\varepsilon_{\mathbf{k}\alpha}A^j+\frac{1}{2}(\partial_i\partial_j\varepsilon_{\mathbf{k}\alpha})A^iA^j]\mathbb{T}^\alpha.
\end{eqnarray}
The third term, which is second order in electromagnetic field, is responsible for the Higgs generation in the conventional theory. This term vanishes when the electron has linear dispersion or flat band.

The geometric contribution to the Higgs mode comes from the pairing terms involving $\tilde{\Delta}_\mathbf{k}(\mathbf{A})=\mathcal{G}^\dagger_{\mathbf{k}-\mathbf{A}}\Delta\mathcal{G}_{\mathbf{k}+\mathbf{A}}$, which are the off-diagonal blocks of the BdG Hamiltonian. As discussed in the main text, this can be expanded in terms of the generators of su(N) as
\begin{eqnarray}
\label{gap}
\tilde{\Delta}_\mathbf{k}(\mathbf{A})=\mathcal{G}^\dagger_{\mathbf{k}-\mathbf{A}}\Delta\mathcal{G}_{\mathbf{k}+\mathbf{A}}=\tilde{\Delta}_{\mathbf{k}\alpha}(\mathbf{A})\mathbb{T}^\alpha.
\end{eqnarray}

To see the appearance of band geometric quantities, we expand the components of $\tilde{\Delta}_\mathbf{k}(\mathbf{A})$ in powers of the external field $\mathbf{A}$. This comes from the expansion of $\mathcal{G}^{(\dagger)}_{\mathbf{k}\pm\mathbf{A}}$ in \eqref{Gexpand}. We further separate the time-dependent Higgs mode $\delta\Delta_\alpha(t)$ in the coefficient of \eqref{gap}: $\tilde{\Delta}_{\mathbf{k}\alpha}(\mathbf{A})=\Delta_0\mathbb{1}+\delta\Delta_\alpha(t)\mathbb{T}^\alpha$.

For $\alpha=0$, we have
\begin{eqnarray}
\label{Delta0}
\tilde{\Delta}_{\mathbf{k}0}(\mathbf{A})&=&\Delta_0-2i\delta\Delta_\alpha\mathcal{A}_{\mathbf{k}i}^\alpha A^i-4\Delta_0\mathcal{A}_{\mathbf{k}i\alpha}\mathcal{A}_{\mathbf{k}j}^\alpha A^iA^j;
\end{eqnarray}
while for $\alpha>0$, we have
\begin{eqnarray}
\label{DeltaAlpha}
\tilde{\Delta}_{\mathbf{k}\alpha}(\mathbf{A})=\delta\Delta_\alpha-2i\Delta_0\mathcal{A}_{\mathbf{k}i\alpha}A^i-4\Delta_0g_{\mathbf{k},ij\alpha}A^iA^j-2ih_{\beta\gamma\alpha}\delta\Delta^\beta\mathcal{A}_{\mathbf{k}i}^\gamma A^i-4\Delta_0h_{\beta\gamma\alpha}\mathcal{A}_{\mathbf{k}i}^\beta\mathcal{A}_{\mathbf{k}j}^\gamma A^iA^j.\nonumber
\end{eqnarray}

Here, $\mathcal{A}_\mathbf{k}\equiv i\mathcal{G}^\dagger_\mathbf{k}\nabla_\mathbf{k}\mathcal{G}_\mathbf{k}=\mathcal{A}_{\mathbf{k}\alpha}\mathbb{T}^\alpha$ is the Berry connection in matrix form and $g_{\mathbf{k},ij}=g_{\mathbf{k},ij\alpha}\mathbb{T}^\alpha$ is the quantum metric. These enters into the pseudomagnetic field, which in turn, enters into the equations of motion for the pseudospins as discussed in the main text.

\section{Gapped graphene.} 
\label{gappedgraphene}
The monolayer graphene with different on-site potentials is described by the Hamiltonian
\begin{eqnarray}
H&=&\sum_\mathbf{k}\Psi^\dagger_\mathbf{k}
\begin{pmatrix}
\varepsilon_A & t\sum_{i=1}^3e^{i\mathbf{k}\cdot\mathbf{\delta}_i}\\
t\sum_{i=1}^3e^{-i\mathbf{k}\cdot\mathbf{\delta}_i} & \varepsilon_B
\end{pmatrix}
\Psi_\mathbf{k},
\end{eqnarray}
where $\Psi^\dagger_\mathbf{k}\equiv(c^\dagger_{A\mathbf{k}},c^\dagger_{B\mathbf{k}})$ and the operator $c^\dagger_{A\mathbf{k}}$ ($c^\dagger_{B\mathbf{k}}$) creates an electron at sublattice A (B). The bond vectors are given by $\mathbf{\delta}_1=\frac{1}{2}(1,\sqrt{3})$, $\mathbf{\delta}_2=\frac{1}{2}(1,-\sqrt{3})$, and $\mathbf{\delta}_3=(-1,0)$. 

The first-order equations of motion are then
\begin{eqnarray}
\label{linearx}
\partial_t\delta\sigma^{x(1)}_\mathbf{k}&=&4\Delta_0\sigma^z_0\mathcal{A}_{\mathbf{k}j}A^j(t)-2\varepsilon_\mathbf{k}\delta\sigma^{y(1)}_\mathbf{k}\\
\label{lineary}
\partial_t\delta\sigma^{y(1)}_\mathbf{k}&=&2\Delta_0\delta\sigma^{z(1)}_\mathbf{k}+2\varepsilon_\mathbf{k}\delta\sigma^{x(1)}_\mathbf{k}+2\sigma^z_0\delta\Delta^{(1)}\\
\label{linearz}
\partial_t\delta\sigma^{z(1)}_\mathbf{k}&=&-4\Delta_0\mathcal{A}_{\mathbf{k}j}A^j(t)\sigma^x_0-2\Delta_0\delta\sigma^{y(1)}_\mathbf{k}.
\end{eqnarray}

From \eqref{linearx} and \eqref{linearz}, along with the intial conditions $\delta\sigma^{x(1)}_\mathbf{k}(0)=\delta\sigma^{y(1)}_\mathbf{k}(0)=0$, one can show that $\Delta_0\delta\sigma^{x(1)}_\mathbf{k}(t)=\varepsilon_\mathbf{k}\delta\sigma^{z(1)}_\mathbf{k}(t)$ at all times. Hence, we can eliminate $\delta\sigma^{z(1)}_\mathbf{k}$ and reduce the number of equations. The initial conditions for zeroth order are $\sigma^x_\mathbf{k}(0)=\Delta_0/\omega_\mathbf{k}$, $\sigma^z_\mathbf{k}(0)=-\varepsilon_\mathbf{k}/\omega_\mathbf{k}$, and $\sigma^y_\mathbf{k}(0)=0$ with $\omega_\mathbf{k}=2\sqrt{\varepsilon_\mathbf{k}^2+\Delta_0^2}$.

The solution in Laplace space is
\begin{eqnarray}
\label{sigmax1}
\tilde{\delta}\sigma^{x(1)}_\mathbf{k}(s)&=&\frac{4\sigma^z_0}{s^2+\omega_\mathbf{k}^2}\big[\sigma^z_0s\mathcal{A}_{\mathbf{k}j}\tilde{A}^j(s)-\varepsilon_\mathbf{k}\tilde{\delta}\Delta^{(1)}(s)\big]\\
\label{sigmay1}
\tilde{\delta}\sigma^{y(1)}_\mathbf{k}(s)&=&\frac{2\sigma^z_0}{s^2+\omega_\mathbf{k}^2}\big[\frac{\omega_\mathbf{k}^2\Delta_0}{\varepsilon_\mathbf{k}}\mathcal{A}_{\mathbf{k}j}\tilde{A}^j(s)+s\tilde{\delta}\Delta^{(1)}(s)\big].
\end{eqnarray}

We note that here the momentum is measured with respect to the valley $\mathbf{K}$. Let us consider when the Cooper pairs have center-of-mass momentum $\mathbf{p}$. This can be realized experimentally by inducing a supercurrent on the superconductor. We can write the momentum of an electron (half of the pair) as $\mathbf{k}=\mathbf{k}_F+\mathbf{p}/2$ where $\mathbf{k}_F$ is a Fermi momentum. We will eventually sum $\mathbf{k}_F$ over the Fermi surface. We assume that $p\ll k_F$ so that we can expand:
\begin{align}
\sigma^z_{0,\mathbf{k}_F+\mathbf{p}/2}&\approx \frac{\mathbf{p}}{2}\cdot\nabla_F\sigma^z_{0,\mathbf{k}_F}=-\frac{\mathbf{p}\cdot \mathbf{v}_F}{4\Delta_0},\;\;
\mathcal{A}_{\mathbf{k}_F+\mathbf{p}/2,j}\approx \mathcal{A}_{\mathbf{k}_F,j}+\frac{\mathbf{p}}{2}\cdot\nabla_F\mathcal{A}_{\mathbf{k}_F,j}\\
\varepsilon_{\mathbf{k}_F+\mathbf{p}/2}&\approx \frac{1}{2}\mathbf{p}\cdot\mathbf{v}_F,\;\;
\frac{1}{s^2+\omega_{\mathbf{k}_F+\mathbf{p}/2}^2}\approx\frac{1}{s^2+4\Delta_0^2}\left(1-\frac{\Delta_0\mathbf{p}\cdot\nabla_F\omega_{\mathbf{k}_F}}{s^2+4\Delta_0^4}\right)
\end{align}
where $\nabla_F$ mean derivative with respect to $\mathbf{k}_F$ and $\mathbf{v}_F=\nabla_F\varepsilon_{\mathbf{k}_F}$ is the Fermi velocity.

Recall that the energy is measured with respect to the Fermi level so that $\varepsilon_{\mathbf{k}_F}=0$. It follows that $\nabla_F\omega_{\mathbf{k}_F}=0$. Eq.\eqref{sigmax1} and \eqref{sigmay1} become
\begin{align}
\tilde{\delta}\sigma^{x(1)}_{\mathbf{k}_F+\mathbf{p}}&=\frac{2\Delta_0s\tilde{A}^j(s)\mathcal{A}_{\mathbf{k}_F,j}}{s^2+4\Delta_0^2}(\mathbf{p}\cdot\nabla_F\sigma^z_{0,\mathbf{k}_F})\\
\tilde{\delta}\sigma^{y(1)}_{\mathbf{k}_F+\mathbf{p}}&=\frac{1}{s^2+4\Delta_0^2}\big[-4\Delta_0^2\mathcal{A}_{\mathbf{k}_F,j}\tilde{A}^j(s)-2\Delta_0^2\tilde{A}^j(s)\mathbf{p}\cdot\nabla_F\mathcal{A}_{\mathbf{k}_F,j}+s\tilde{\delta}\Delta^{(1)}\mathbf{p}\cdot\nabla_F\sigma^z_{0,\mathbf{k}_F}\big]
\end{align}

We now sum over the Fermi surface $\sum_{\mathbf{k}_F}$ using the approximation $\vec{\mathcal{A}}_{-\mathbf{k}_F}=-\vec{\mathcal{A}}_{\mathbf{k}_F}$ which is valid so long as the chemical potential is not so large so that the massive Dirac Hamiltonian is a good description for each valleys. We obtain
\begin{eqnarray}
\sum_{\mathbf{k}_F}\tilde{\delta}\sigma^{x(1)}_{\mathbf{k}_F+\mathbf{p}}&=&-\sum_{\mathbf{k}_F}\frac{s(\mathbf{p}\cdot\mathbf{v}_F)(\mathcal{A}_{\mathbf{k}_F}\cdot\mathbf{A})}{(s^2+4\Delta_0^2)(s+i\Omega)}\\
\sum_{\mathbf{k}_F}\tilde{\delta}\sigma^{y(1)}_{\mathbf{k}_F+\mathbf{p}}&=&-\frac{2\Delta_0^2\tilde{A}^j(s)}{s^2+4\Delta_0^2}\mathbf{p}\cdot\sum_{\mathbf{k}_F}\nabla_F\mathcal{A}_{\mathbf{k}_Fj}.
\end{eqnarray}

This gives the Higgs mode in Laplace space now written as
\begin{eqnarray}
\tilde{\delta}\Delta^{(1)}_\mathbf{p}(s)=U\sum_{\mathbf{k}_F}[\tilde{\delta}\sigma^{x(1)}_{\mathbf{k}_F+\mathbf{p}}(s)+i\tilde{\delta}\sigma^{y(1)}_{\mathbf{k}_F+\mathbf{p}}(s)].
\end{eqnarray}

To calculate the second order correction to the Higgs mode, we need the explicit first-order solutions of the pseudospins $\delta\sigma^{x(1)}_\mathbf{k}(t)$ and $\delta\sigma^{x(1)}_\mathbf{k}(t)$. We define $\mathbb{B}\equiv \sum_{\mathbf{k}_F}\nabla_F\mathcal{A}_{\mathbf{k}_F}$ and $\mathbb{C}\equiv\sum_{\mathbf{k}_F}\mathbf{v}_F\mathcal{A}_{\mathbf{k}_F}$.
They are given by:
\begin{align}
\delta\sigma^{x(1)}_\mathbf{k}(t)=&C_{\mathbf{k}1}e^{-i\omega_\mathbf{k}t}+C_{\mathbf{k}2}e^{i\omega_\mathbf{k}t}+C_{\mathbf{k}3}e^{-i2\Delta_0t}+C_{\mathbf{k}4}e^{i2\Delta_0t}+C_{\mathbf{k}5}e^{-i\Omega t}\\
\delta\sigma^{y(1)}_\mathbf{k}(t)=&D_{\mathbf{k}1}e^{-i\omega_\mathbf{k}t}+D_{\mathbf{k}2}e^{i\omega_\mathbf{k}t}+D_{\mathbf{k}3}e^{-i2\Delta_0t}+D_{\mathbf{k}4}e^{i2\Delta_0t}+D_{\mathbf{k}5}e^{-i\Omega t}
\end{align}
where
\begin{eqnarray}
C_{\mathbf{k}1}&=&-\frac{2i\Delta_0\sigma^z_0(\mathcal{A}_\mathbf{k}\cdot\mathbf{A})}{\Omega-\omega_\mathbf{k}}-\frac{2i\varepsilon_\mathbf{k}\sigma^z_0U(\mathbf{p}\cdot\mathbb{C}\cdot\mathbf{A})}{(2\Delta_0-\omega_\mathbf{k})(2\Delta_0+\omega_\mathbf{k})(\Omega-\omega_\mathbf{k})}+\frac{4i\varepsilon_\mathbf{k}\sigma^z_0\Delta_0^2U(\mathbf{p}\cdot\mathbb{B}\cdot\mathbf{A})}{(2\Delta_0-\omega_\mathbf{k})\omega_\mathbf{k}(2\Delta_0+\omega_\mathbf{k})(\Omega-\omega_\mathbf{k})}\\
C_{\mathbf{k}2}&=&\frac{2i\Delta_0\sigma^z_0(\mathcal{A}_\mathbf{k}\cdot\mathbf{A})}{\Omega+\omega_\mathbf{k}}
+\frac{2i\varepsilon_\mathbf{k}\sigma^z_0U(\mathbf{p}\cdot\mathbb{C}\cdot\mathbf{A})}{(2\Delta_0-\omega_\mathbf{k})(2\Delta_0+\omega_\mathbf{k})(\Omega+\omega_\mathbf{k})}-\frac{4i\varepsilon_\mathbf{k}\sigma^z_0\Delta_0^2U(\mathbf{p}\cdot\mathbb{B}\cdot\mathbf{A})}{(2\Delta_0-\omega_\mathbf{k})\omega_\mathbf{k}(2\Delta_0+\omega_\mathbf{k})(\Omega+\omega_\mathbf{k})}\\
C_{\mathbf{k}3}&=&\frac{2i\varepsilon_\mathbf{k}\sigma^z_0U(\mathbf{p}\cdot\mathbb{C}\cdot\mathbf{A})}{(2\Delta_0-\omega_\mathbf{k})(2\Delta_0+\omega_\mathbf{k})(\Omega-2\Delta_0)}
-\frac{2i\varepsilon_\mathbf{k}\sigma^z_0\Delta_0U(\mathbf{p}\cdot\mathbb{B}\cdot\mathbf{A})}{(2\Delta_0-\omega_\mathbf{k})(2\Delta_0+\omega_\mathbf{k})(\Omega-2\Delta_0)}\\
C_{\mathbf{k}4}&=&\frac{2i\varepsilon_\mathbf{k}\sigma^z_0U(\mathbf{p}\cdot\mathbb{C}\cdot\mathbf{A})}{(2\Delta_0-\omega_\mathbf{k})(2\Delta_0+\omega_\mathbf{k})(\Omega+2\Delta_0)}
+\frac{2i\varepsilon_\mathbf{k}\sigma^z_0\Delta_0U(\mathbf{p}\cdot\mathbb{B}\cdot\mathbf{A})}{(2\Delta_0-\omega_\mathbf{k})(2\Delta_0+\omega_\mathbf{k})(\Omega+2\Delta_0)}\\
C_{\mathbf{k}5}&=&\frac{4i\Omega\Delta_0\sigma^z_0(\mathcal{A}_\mathbf{k}\cdot\mathbf{A})}{(\Omega-\omega_\mathbf{k})(\Omega+\omega_\mathbf{k})}
-\frac{4i\Omega\varepsilon_\mathbf{k}\sigma^z_0U(\mathbf{p}\cdot\mathbb{C}\cdot\mathbf{A})}{(\Omega-2\Delta_0)(\Omega+2\Delta_0)(\Omega-\omega_\mathbf{k})(\Omega+\omega_\mathbf{k})}
+\frac{8i\varepsilon_\mathbf{k}\sigma^z_0\Delta_0^2U(\mathbf{p}\cdot\mathbb{B}\cdot\mathbf{A})}{(\Omega-2\Delta_0)(\Omega+2\Delta_0)(\Omega-\omega_\mathbf{k})(\Omega+\omega_\mathbf{k})}
\end{eqnarray}
and
\begin{eqnarray}
D_{\mathbf{k}1}&=&-\frac{\Delta_0(\mathcal{A}\cdot\mathbf{A})}{\Omega-\omega_\mathbf{k}}+\frac{\sigma^z_0U(\mathbf{p}\cdot\mathbb{C}\cdot\mathbf{A})\omega_\mathbf{k}}{(2\Delta_0-\omega_\mathbf{k})(2\Delta_0+\omega_\mathbf{k})(\Omega-\omega_\mathbf{k})}-\frac{2\Delta_0^2U\sigma^z_0(\mathbf{p}\cdot\mathbb{B}\cdot\mathbf{A})}{(2\Delta_0-\omega_\mathbf{k})(2\Delta_0+\omega_\mathbf{k})(\Omega-\omega_\mathbf{k})}\\
D_{\mathbf{k}2}&=&\frac{\Delta_0(\mathcal{A}\cdot\mathbf{A})}{\Omega+\omega_\mathbf{k}}-\frac{\sigma^z_0U(\mathbf{p}\cdot\mathbb{C}\cdot\mathbf{A})_\mathbf{k}}{(2\Delta_0-\omega_\mathbf{k})(2\Delta_0+\omega_\mathbf{k})(\Omega+\omega_\mathbf{k})}+\frac{2\Delta_0^2U\sigma^z_0(\mathbf{p}\cdot\mathbb{B}\cdot\mathbf{A})}{(2\Delta_0-\omega_\mathbf{k})(2\Delta_0+\omega_\mathbf{k})(\Omega+\omega_\mathbf{k})}\\
D_{\mathbf{k}3}&=&-\frac{2\sigma^z_0\Delta_0U(\mathbf{p}\cdot\mathbb{C}\cdot\mathbf{A})}{(2\Delta_0-\omega_\mathbf{k})(2\Delta_0+\omega_\mathbf{k})(\Omega-2\Delta_0)}+\frac{2\Delta_0^2U\sigma^z_0(\mathbf{p}\cdot\mathbb{B}\cdot\mathbf{A})}{(2\Delta_0-\omega_\mathbf{k})(2\Delta_0+\omega_\mathbf{k})(\Omega-2\Delta_0)}\\
D_{\mathbf{k}4}&=&\frac{2\sigma^z_0\Delta_0U(\mathbf{p}\cdot\mathbb{C}\cdot\mathbf{A})}{(2\Delta_0-\omega_\mathbf{k})(2\Delta_0+\omega_\mathbf{k})(\Omega+2\Delta_0)}+\frac{2\Delta_0^2U\sigma^z_0(\mathbf{p}\cdot\mathbb{B}\cdot\mathbf{A})}{(2\Delta_0-\omega_\mathbf{k})(2\Delta_0+\omega_\mathbf{k})(\Omega+2\Delta_0)}\\
D_{\mathbf{k}5}&=&\frac{2\Delta_0\omega_\mathbf{k}(\mathcal{A}\cdot\mathbf{A})}{(\Omega-\omega_\mathbf{k})(\Omega+\omega_\mathbf{k})}+\frac{2\sigma^z_0U(\mathbf{p}\cdot\mathbb{C}\cdot\mathbf{A})\Omega^2}{(\Omega-2\Delta_0)(\Omega+2\Delta_0)(\Omega-\omega_\mathbf{k})(\Omega+\omega_\mathbf{k})}-\frac{4\Delta_0^2U\sigma^z_0\Omega(\mathbf{p}\cdot\mathbb{B}\cdot\mathbf{A})}{(\Omega-2\Delta_0)(\Omega+2\Delta_0)(\Omega-\omega_\mathbf{k})(\Omega+\omega_\mathbf{k})}.
\end{eqnarray}

\subsection{Second order calculations.} 
The second order equations of motion are
\begin{align}
\partial_t\delta\sigma^{x(2)}_\mathbf{k}=&-2\varepsilon_\mathbf{k}\delta\sigma^{y(2)}_\mathbf{k}+4\Delta_0\mathcal{A}_{\mathbf{k}j}A^j\delta\sigma^{z(1)}_0+4\delta\Delta^{(1)}\mathcal{A}_{\mathbf{k}j}A^j\sigma^z_0\\
\partial_t\delta\sigma^{y(2)}_\mathbf{k}=&-8\Delta_0g_{\mathbf{k},ij}A^iA^j\sigma^z_0-8\Delta_0\mathcal{A}_{\mathbf{k}i}\mathcal{A}_{\mathbf{k}j}A^iA^j\sigma^z_0+2\Delta_0\delta\sigma^{z(2)}_\mathbf{k}+2\varepsilon_\mathbf{k}\delta\sigma^{x(2)}_\mathbf{k}+(\partial_i\partial_j\varepsilon_\mathbf{k})A^iA^j\sigma^x_0\nonumber\\
&+2\sigma^z_0\delta\Delta^{(2)}+2\delta\Delta^{(1)}\delta\sigma^{z(1)}_\mathbf{k}\\
\partial_t\delta\sigma^{z(2)}_\mathbf{k}=&-2\Delta_0\delta\sigma^{y(2)}_\mathbf{k}-2\delta\Delta^{(1)}\delta\sigma^{y(1)}_\mathbf{k}-4\Delta_0\mathcal{A}_{\mathbf{k}j}A^j\delta\sigma^{x(1)}_\mathbf{k}-4\delta\Delta^{(1)}\mathcal{A}_{\mathbf{k}j}A^j\sigma^x_0.
\end{align}

We perform Laplace transform to the equations above. We only need $\tilde{\delta}\sigma^{x(2)}_\mathbf{k}$ and $\tilde{\delta}\sigma^{y(2)}_\mathbf{k}$ given by
\begin{align}
\label{secondorder1}
\tilde{\delta}\sigma^{x(2)}_\mathbf{k}=&\frac{(s^2+4\Delta_0^2)F_{\mathbf{k}1}(s)-s(\mathbf{p}\cdot\mathbf{v}_F)F_{\mathbf{k}2}(s)}{s^3+4\Delta_0^2s+2s\varepsilon_\mathbf{k}(\mathbf{p}\cdot\mathbf{v}_F)}-\frac{2\Delta_0(\mathbf{p}\cdot\mathbf{v}_F)F_{\mathbf{k}3}(s)}{s^3+4\Delta_0^2s+2s\varepsilon_\mathbf{k}(\mathbf{p}\cdot\mathbf{v}_F)}\\
\label{secondorder2}
\tilde{\delta}\sigma^{y(2)}_\mathbf{k}=&\frac{2s\varepsilon_\mathbf{k}F_{\mathbf{k}1}(s)+s^2F_{\mathbf{k}2}(s)+2s\Delta_0F_{\mathbf{k}3}(s)}{s^3+4\Delta_0^2s+2s\varepsilon_\mathbf{k}(\mathbf{p}\cdot\mathbf{v}_F)},
\end{align}
where
\begin{eqnarray}
F_{\mathbf{k}1}(s)&=&\frac{4\Delta_0^2}{\varepsilon_\mathbf{k}}(\mathcal{A}\cdot\mathbf{A})\left[\frac{C_{\mathbf{k}1}}{s+i(\Omega+\omega_\mathbf{k})}+\frac{C_{\mathbf{k}2}}{s+i(\Omega-\omega_\mathbf{k})}+\frac{C_{\mathbf{k}3}}{s+i(\Omega+2\Delta_0)}+\frac{C_{\mathbf{k}4}}{s+i(\Omega-2\Delta_0)}+\frac{C_{\mathbf{k}5}}{s+i2\Omega}\right]\nonumber\\
&{}&{}-\mathbf{p}\cdot(\mathbb{C}+\Delta_0\mathbb{B})\cdot\mathbf{A}\frac{2iU\sigma^z_0(\mathcal{A}_\mathbf{k}\cdot\mathbf{A})}{(\Omega-2\Delta_0)[s+i(\Omega+2\Delta_0)]}+\mathbf{p}\cdot(\mathbb{C}+\Delta_0\mathbb{B})\cdot\mathbf{A}\frac{2iU\sigma^z_0(\mathcal{A}_\mathbf{k}\cdot\mathbf{A})}{(\Omega+2\Delta_0)[s+i(\Omega-2\Delta_0)]}\\
&{}&{}-\mathbf{p}\cdot(\Omega\mathbb{C}+2\Delta_0^2\mathbb{B})\cdot\mathbf{A}\frac{4iU\sigma^z_0(\mathcal{A}_\mathbf{k}\cdot\mathbf{A})}{(\Omega-2\Delta_0)(\Omega+2\Delta_0)}\\
F_{\mathbf{k}2}(s)&=&-8\Delta_0(g_{\mathbf{k},ij}+\mathcal{A}_{\mathbf{k}i}\mathcal{A}_{\mathbf{k}j}\sigma^z_0)\frac{A^iA^j}{s+2i\Omega}+(\partial_i\partial_j\varepsilon_\mathbf{k})\sigma^x_{\mathbf{k},0}\frac{A^iA^j}{s+2i\Omega}+2\sigma^z_{\mathbf{k},0}\Tilde{\delta}\Delta^{(2)}\\
\label{f3}
F_{\mathbf{k}3}(s)&=&-2\mathcal{L}\big\{\delta\Delta^{(1)}_\mathbf{p}(t)\delta\sigma^{y(1)}_\mathbf{k}\big\}-4\Delta_0(\mathcal{A}\cdot\mathbf{A})\mathcal{L}\big\{e^{-i\Omega t}\delta\sigma^{x(1)}_\mathbf{k}(t)\big\}-4\sigma^x_{\mathbf{k},0}(\mathcal{A}\cdot\mathbf{A})\mathcal{L}\big\{e^{-i\Omega t}\delta\Delta^{(1)}\big\}.
\end{eqnarray}

The Higgs mode in Laplace space is given by
\begin{eqnarray}
\label{higgs2}
\tilde{\delta}\Delta^{(2)}_\mathbf{p}=U\sum_{\mathbf{k}_F}(\tilde{\delta}\sigma^{x(2)}_{\mathbf{k}_F+\mathbf{p}/2}+i\tilde{\delta}\sigma^{y(2)}_{\mathbf{k}_F+\mathbf{p}/2}).
\end{eqnarray}

In calculating the second order equations of motion, we need the following Laplace transforms appearing in \eqref{f3}:
\begin{eqnarray}
\mathcal{L}\big\{\delta\Delta^{(1)}_\mathbf{p}(t)\delta\sigma^{y(1)}_\mathbf{k}\big\}&=&(D_{\mathbf{k}4}E_{\mathbf{k}1}+D_{\mathbf{k}3}E_{\mathbf{k}2})\frac{1}{s}+\frac{D_{\mathbf{k}4}E_{\mathbf{k}2}}{s-4\Delta_0i}+\frac{D_{\mathbf{k}3}E_{\mathbf{k}1}}{s+4\Delta_0i}+\frac{D_{\mathbf{k}2}E_{\mathbf{k}1}}{s+i(2\Delta_0-\omega_\mathbf{k})}+\frac{D_{\mathbf{k}1}E_{\mathbf{k}2}}{s-i(2\Delta_0-\omega_\mathbf{k})}\nonumber\\
&{}&{}+\frac{D_{\mathbf{k}2}E_{\mathbf{k}2}}{s-i(2\Delta_0+\omega_\mathbf{k})}+\frac{D_{\mathbf{k}1}E_{\mathbf{k}1}}{s+i(2\Delta_0+\omega_\mathbf{k})}+\frac{D_{\mathbf{k}2}E_{\mathbf{k}3}}{s+i(\Omega-\omega_\mathbf{k})}+\frac{D_{\mathbf{k}5}E_{\mathbf{k}2}}{s+i(\Omega-2\Delta_0)}\nonumber\\
&{}&{}+\frac{D_{\mathbf{k}4}E_{\mathbf{k}3}}{s+i(\Omega-2\Delta_0)}+\frac{D_{\mathbf{k}5}E_{\mathbf{k}3}}{s+i2\Omega}+\frac{D_{\mathbf{k}5}E_{\mathbf{k}1}+D_{\mathbf{k}3}E_{\mathbf{k}3}}{s+i(\Omega+2\Delta_0)}+\frac{D_{\mathbf{k}1}E_{\mathbf{k}3}}{s+i(\Omega+\omega_\mathbf{k})}\\
\mathcal{L}\big\{e^{-i\Omega t}\delta\sigma^{x(1)}_\mathbf{k}(t)\big\}&=&\frac{C_{\mathbf{k}1}}{s+i(\Omega+\omega_\mathbf{k})}+\frac{C_{\mathbf{k}2}}{s+i(\Omega-\omega_\mathbf{k})}+\frac{C_{\mathbf{k}3}}{s+i(\Omega+2\Delta_0)}+\frac{C_{\mathbf{k}4}}{s+i(\Omega-2\Delta_0)}+\frac{C_{\mathbf{k}5}}{s+i2\Omega}\\
\mathcal{L}\big\{e^{-i\Omega t}\delta\Delta^{(1)}\big\}&=&\frac{E_{\mathbf{k}1}}{s+i(\Omega+2\Delta_0)}+\frac{E_{\mathbf{k}2}}{s+i(\Omega-2\Delta_0)}+\frac{E_{\mathbf{k}3}}{s+i2\Omega}.
\end{eqnarray}

Here the constants $E_{\mathbf{k}1}$, $E_{\mathbf{k}2}$, and $E_{\mathbf{k}3}$ are given by
\begin{align}
E_{\mathbf{k}1}=-\frac{iU\mathbf{p}\cdot(\mathbb{C}+\Delta_0\mathbb{B})\cdot\mathbf{A}}{2(\Omega-2\Delta_0)},\;\;
E_{\mathbf{k}2}=\frac{iU\mathbf{p}\cdot(\mathbb{C}+\Delta_0\mathbb{B})\cdot\mathbf{A}}{2(\Omega+2\Delta_0)},\;\;
E_{\mathbf{k}3}=-\frac{iU\mathbf{p}\cdot(\Omega\mathbb{C}+2\Delta_0^2\mathbb{B})\cdot\mathbf{A}}{(\Omega-2\Delta_0)(\Omega+2\Delta_0)}.
\end{align}

Substituting \eqref{secondorder1} and \eqref{secondorder2} into \eqref{higgs2} and performing an inverse Laplace transform, we get
\begin{eqnarray}
\label{deltasecond2}
\delta\Delta^{(2)}_\mathbf{p}(t)=[\mathbb{C}_{band}^{(2)}(t)_{ij}+\mathbb{B}_{geom}^{(2)}(t,\mathcal{A},g)_{ij}]A^iA^j.
\end{eqnarray}

As described in the main text, we divided the result into two main contributions. The conventional term $\mathbb{C}_{band}^{(2)}(t)_{ij}$ is dependent on band curvature $\partial_i\partial_j\varepsilon_{\mathbf{k}_F}$ and has the form
\begin{eqnarray}
\mathbb{C}_{band}^{(2)}(t)_{ij}
&=&
\frac{iU}{4}\sum_{\mathbf{k}_F}\big[\kappa_1(t)+\kappa_2(t)\mathbf{p}\cdot\mathbf{v}_{\mathbf{k}_F}\big]\partial_i\partial_j\varepsilon_{\mathbf{k}_F}.
\end{eqnarray}
Here, the second term inside the square brackets contribute when there is supercurrent $\mathbf{p}\neq 0$. The functions $\kappa_1(t)$ and $\kappa_2(t)$ contain the explicit time dependence of the oscillation
\begin{align}
\kappa_1(t)=i\left[\frac{\Omega e^{-i2\Omega t}}{\Omega^2-\Delta_0^2}-\frac{e^{-i2\Delta_0t}}{2(\Omega-\Delta_0)}-\frac{e^{i2\Delta_0t}}{2(\Omega+\Delta_0)}\right],\;\;
\kappa_2(t)=\frac{1}{2}\left[\frac{e^{-i2\Omega t}}{\Omega^2-\Delta_0^2}-\frac{e^{-i2\Delta_0t}}{2\Delta_0(\Omega-\Delta_0)}+\frac{e^{i2\Delta_0t}}{2\Delta_0(\Omega+\Delta_0)}\right].
\end{align}

Note that the denominators contain the information about the Anderson pseudospin resonance $\Omega=\pm\Delta_0$, with the $\pm$ coming from the particle-hole symmetry.

As discussed in the main text, the presence of the supercurrent allows a linear coupling with the external field. In the second order, the presence of $\mathbf{p}\neq 0$ does not lead to interesting physics and only vastly complicates the explicit expression for the geometric contribution. Hence, we only confined to the $\mathbf{p}=0$ case. The geometric part is now given by
\begin{align}
\mathbb{B}_{geom}^{(2)}(t,\mathcal{A},g)_{ij}=-4\Delta_0 U\sum_{\mathbf{k}_F}\big[\beta_1(t)\mathcal{A}_{\mathbf{k}_F,i}\mathcal{A}_{\mathbf{k}_F,j}+\beta_1(t)g_{\mathbf{k}_F,ij}\big].
\end{align}
Note the appearance of the geometric quantities $\mathcal{A}_{\mathbf{k}_F,i}\mathcal{A}_{\mathbf{k}_F,j}$ and $g_{\mathbf{k}_F,ij}$.

The functions $\beta_1(t)$ and $\beta_2(t)$ contain the explicit time dependence
\begin{align}
\beta_1(t)=\frac{\Delta_0}{\Omega^2-4\Delta_0^2}\big[e^{-i(\Omega+2\Delta_0)t}+e^{-i(\Omega-2\Delta_0)t}+e^{-i2\Omega t}-3\big],\;\;
\beta_2(t)=i\left[\frac{\Omega e^{-i2\Omega t}}{\Omega^2-\Delta_0^2}-\frac{e^{-i2\Delta_0t}}{2(\Omega-\Delta_0)}-\frac{e^{i2\Delta_0t}}{2(\Omega+\Delta_0)}\right].
\end{align}

Note that the denominators contain the known Higgs gap $\Omega=\pm 2\Delta_0$ and the Anderson pseudospin resonance $\Omega=\pm\Delta_0$.

\section{Harper-Hubbard model.}
\label{harperhubbard}
Similar method can be applied to calculate the Higgs mode for this model. One major difference from the previous example is the fact that now we have to integrate over the entire Brillouin zone since now the band is practically flat and there is no well defined Fermi surface. Fortunately, the independence of the energy dispersion to $\mathbf{k}$ allows us to perform the integrals easily. The first order Higgs mode vanishes as a consequence of the symmetry $\mathcal{A}_{-\mathbf{k}}=-\mathcal{A}_{\mathbf{k}}$. Explicitly, the Higgs mode in Laplace space is
\begin{eqnarray}
\label{deltaharperhubbard}
\tilde{\Delta}^{(2)}_R(s)&=&\frac{\alpha_1(s)G_{ij}A^iA^j+\alpha_2(s)B_{ij}A^iA^j}{(s^2+\Omega_1^2)(s^2+\Omega_2^2)(s^2+\Omega_3^2)(s^2+\Omega_4^2)}.
\end{eqnarray}

The important physics of the denominator and the quantum metric factor was discussed in the main text. Here we report the detailed form of the functions $\alpha_1(s)$ and $\alpha_1(s)$ that were left out in the main text:
\begin{align}
\alpha_1(s)=&\varepsilon^2\big[s^2+(\omega-\Omega)^2\big]\big[s^2+(\omega+\Omega)^2\big]\omega(s^2+\omega^2)16\lambda\Delta_0\Omega^2(s^2+\omega^2-4\pi^2\lambda\omega)\nonumber\\
&+16\lambda^2\varepsilon^2\Delta_0\Omega^2(2\pi)^2s^2\big[s^2+(\omega-\Omega)^2\big]\big[s^2+(\omega+\Omega)^2\big]\\
\alpha_2(s)=&[\varepsilon^2s^4-3\Delta^2_0s^3+2\varepsilon^2(\omega^2+\Omega^2)s^2-\Delta_0^2(\omega^2+3\Omega^2)s+\varepsilon^2(\omega^2-\Omega^2)^2]\omega(s^2+\omega^2) 16\lambda\Delta_0\Omega^2(s^2+\omega^2-(2\pi)^2\lambda\omega)\nonumber\\&
+16\lambda^2\varepsilon^2\Delta_0\Omega^2(2\pi)^2s^2\big[s^2+(\omega-\Omega)^2\big]\big[s^2+(\omega+\Omega)^2\big],
\end{align}
where $\lambda=U/(2\pi)^2$.\\

\section{Twisted bilayer graphene.} 
\label{tbg}
As discussed in the main text, we used the ten-band model which captures the correct energetics, symmetry, and topology of the active bands \cite{Po2019}. In this model, the $p_\pm$ and $p_z$ orbitals are attached to the triangular lattice; $s$ to the Kagome; and $p_\pm$ to the honeycomb lattice. We solve numerically the Higgs mode using a short external pulse of the form $\mathbf{A}(t)=\mathbf{A}\exp \left(t/\tau\right)^2$ \cite{Chou2017}, with $\tau=\pi\hbar/\Delta_0$. 

The electric field is chosen to be linearly polarized along the x direction. We found similar results for polarizations along $\hat{x}+\hat{y}$ and $\hat{y}$ directions. We consider the case where the electromagnetic pulse is directly incident on the sample and that there is no supercurrent so that the linear contributions discussed in the main text vanish when the contributions from the two valleys are summed. The main geometric contribution then comes from the quantum metric $g_{ij}(\mathbf{k})A^iA^j$.

\subsection{Analytical solution.} 
We assume the general pairing case $\Delta_\mathbf{k}=\Delta_0f_\mathbf{k}=\Delta_0(u_\mathbf{k}+iv_\mathbf{k})$. The s wave case can be obtained by taking $v_\mathbf{k}=0$ and $u_\mathbf{k}=1$, while the $d+id$ pairing by $u_\mathbf{k}=\cos k_x-\cos\left(k_x/2\right)\cos\left(\sqrt{3}k_y/2\right)$ and $v_\mathbf{k}=\sqrt{3}\sin\left(k_x/2\right)\sin\left(\sqrt{3}k_y/2\right)$.  The pseudomagnetic field has the form $\mathbf{B}_\mathbf{k}=\langle-\Delta^R_\mathbf{k}(\mathbf{A}),-\Delta^I_\mathbf{k}(\mathbf{A}),\varepsilon_\mathbf{k}+\frac{1}{2}\partial_i\partial_j\varepsilon_\mathbf{k}A^iA^j\rangle$. Let us expand the x and y components of the pseudomagnetic field explicitly
\begin{align}
\Delta^R_\mathbf{k}(\mathbf{A})=&\Delta_0u_\mathbf{k}+\Delta^{(1)}_R+2\Delta_0v_\mathbf{k}\mathcal{A}_{\mathbf{k}i}A^i+\Delta^{(2)}_R+2\Delta^{(1)}_I\mathcal{A}_{\mathbf{k}i}A^i-2\Delta_0u_\mathbf{k}(g_{\mathbf{k}},ij+\mathcal{A}_{\mathbf{k}i}\mathcal{A}_{\mathbf{k}j})A^iA^j\\
\Delta^I_\mathbf{k}(\mathbf{A})=&\Delta_0v_\mathbf{k}+\Delta^{(1)}_I-2\Delta_0u_\mathbf{k}\mathcal{A}_{\mathbf{k}i}A^i+\Delta^{(2)}_I-2\Delta^{(1)}_R\mathcal{A}_{\mathbf{k}i}A^i+2\Delta_0v_\mathbf{k}(g_{\mathbf{k},ij}+\mathcal{A}_{\mathbf{k}i}\mathcal{A}_{\mathbf{k}j})A^iA^j
\end{align}

The equations of motion is $\partial_t\vec{\sigma}_\mathbf{k}=2\mathbf{B}_\mathbf{k}\cdot\vec{\sigma}_\mathbf{k}$. In the lowest order, this gives
\begin{align}
-2\Delta_0v_\mathbf{k}\tilde{\sigma}^{z(0)}_\mathbf{k}-2\varepsilon_\mathbf{k}\tilde{\sigma}^{y(0)}_\mathbf{k}&=0\\
2\varepsilon_\mathbf{k}\tilde{\sigma}^{x(0)}_\mathbf{k}+2\Delta_0u_\mathbf{k}\tilde{\sigma}^{z(0)}_\mathbf{k}&=0\\
-2\Delta_0u_\mathbf{k}\tilde{\sigma}^{y(0)}_\mathbf{k}+2\Delta_0v_\mathbf{k}\tilde{\sigma}^{x(0)}_\mathbf{k}&=0
\end{align}

This gives the nontrivial solution
\begin{align}
\label{sigmazero}
\vec{\sigma}^{(0)}_\mathbf{k}=\left(\Delta_0\frac{u_\mathbf{k}}{\omega_\mathbf{k}},\Delta_0\frac{v_\mathbf{k}}{\omega_\mathbf{k}},-\frac{\varepsilon_\mathbf{k}}{\omega_\mathbf{k}}\right),
\end{align}
where $\omega_\mathbf{k}=2\sqrt{\varepsilon_\mathbf{k}^2+\Delta_0^2|f_\mathbf{k}|^2}$.
This simply says that the pseudospin is anti-parallel to the pseudomagnetic field. Note that, unlike the s wave case, we now have a non-zero y component. This is due to the fact that the pairing potential has an imaginary part $v_\mathbf{k}$.
 
We now calculate the first order Higgs mode. The equations of motion are
\begin{align}
\partial_t\sigma^{x(1)}_\mathbf{k}&=-\frac{\varepsilon_\mathbf{k}}{\omega_\mathbf{k}}4\Delta_0u_\mathbf{k}\mathcal{A}_{\mathbf{k}i}A^i-2\varepsilon_\mathbf{k}\sigma^{y(1)}_\mathbf{k}+2\frac{\varepsilon_\mathbf{k}}{\omega_\mathbf{k}}\Delta^{(1)}_I-2\Delta_0v_\mathbf{k}\sigma^{z(1)}_\mathbf{k}\\
\partial_t\sigma^{y(1)}_\mathbf{k}&=2\varepsilon_\mathbf{k}\sigma^{x(1)}_\mathbf{k}+(2\Delta^{(1)}_R+4\Delta_0v_\mathbf{k}\mathcal{A}_{\mathbf{k}i}A^i)\sigma^{z(0)}_\mathbf{k}+2\Delta_0u_\mathbf{k}\sigma^{z(1)}_\mathbf{k}\\
\partial_t\sigma^{z(1)}_\mathbf{k}&=
-2(\Delta^{(1)}_R+2\Delta_0v_\mathbf{k}\mathcal{A}_{\mathbf{k}i}A^i)\frac{\Delta_0v_\mathbf{k}}{\omega_\mathbf{k}}-2\Delta_0u_\mathbf{k}\sigma^{y(1)}_\mathbf{k}+(\Delta^{(1)}_I-2\Delta_0u_\mathbf{k}\mathcal{A}_{\mathbf{k}i}A^i)\frac{\Delta_0u_\mathbf{k}}{\omega_\mathbf{k}}+2\Delta_0v_\mathbf{k}\sigma^{x(1)}_\mathbf{k}.
\end{align}
We can solve the equations of motion using Laplace transformation. The self-consistent equation then gives for the real part
\begin{align}
\tilde{\Delta}^{(1)}_R(s)&=-U\sum_{\mathbf{k}}\tilde{\sigma}^{x(1)}_\mathbf{k}\\
\label{delta1R}
&=-4\Delta_0^2U\sum_{\mathbf{k}}\frac{v_\mathbf{k}^2}{\omega_\mathbf{k}}(\tilde{\Delta}^{(1)}_R+2\Delta_0v_\mathbf{k}\mathcal{A}_{\mathbf{k}i}\tilde{A}^i)\frac{1}{s^2+\omega_\mathbf{k}^2}+4\Delta_0^2U\sum_{\mathbf{k}}\frac{v_\mathbf{k}u_\mathbf{k}}{\omega_\mathbf{k}}(\tilde{\Delta}^{(1)}_I-2\Delta_0u_\mathbf{k}\mathcal{A}_{\mathbf{k}i}\tilde{A}^i)\frac{1}{s^2+\omega_\mathbf{k}^2}.
\end{align}

Similarly, the imaginary part is
\begin{align}
\tilde{\Delta}^{(1)}_I(s)&=-U\sum_{\mathbf{k}}\tilde{\sigma}^{y(1)}_\mathbf{k}\\
\label{delta1I}
&=2\tilde{\Delta}^{(1)}_RU\sum_{\mathbf{k}}\frac{v_\mathbf{k}}{\omega_\mathbf{k}}\frac{2\Delta_0^2u_\mathbf{k}}{s^2+\omega_\mathbf{k}^2}-2\tilde{\Delta}^{(1)}_IU\sum_{\mathbf{k}}\frac{u_\mathbf{k}^2}{\omega_\mathbf{k}}\frac{2\Delta_0^2}{s^2+\omega_\mathbf{k}^2}+8\Delta_0^3\sum_{\mathbf{k}}\frac{u_\mathbf{k}}{s^2+\omega_\mathbf{k}^2}\mathcal{A}_{\mathbf{k}i}\tilde{A}^i\frac{|f_\mathbf{k}|^2}{\omega_\mathbf{k}}
\end{align}

We can solve the system of equations \eqref{delta1R} and \eqref{delta1I}. This gives
\begin{align}
\label{deltar}
\tilde{\Delta}^{(1)}_R(s)=\frac{c_{22}\kappa_1-c_{12}\kappa_2}{c_{11}c_{22}-c_{21}c_{12}},\;\;
\tilde{\Delta}^{(1)}_I(s)=\frac{-c_{21}\kappa_1+c_{11}\kappa_2}{c_{11}c_{22}-c_{21}c_{12}}
\end{align}
where
\begin{align}
\label{c11}
c_{11}=1+4\Delta_0^2U\sum_{\mathbf{k}}\frac{v_\mathbf{k}^2}{\omega_\mathbf{k}}\frac{1}{s^2+\omega_\mathbf{k}^2},\;\;
c_{12}=-4\Delta_0^2U\sum_{\mathbf{k}}\frac{u_\mathbf{k}v_\mathbf{k}}{\omega_\mathbf{k}}\frac{1}{s^2+\omega_\mathbf{k}^2},\;\;
c_{21}=-c_{12},\;\;
c_{22}&=-1-4U\sum_{\mathbf{k}}\frac{\Delta_0^2}{s^2+\omega_\mathbf{k}^2}\frac{u_\mathbf{k}^2}{\omega_\mathbf{k}}\\
\label{kappa2}
\kappa_1=-8\Delta_0^3U\sum_{\mathbf{k}}\frac{v_\mathbf{k}}{\omega_\mathbf{k}}\frac{|f_\mathbf{k}|^2\mathcal{A}_{\mathbf{k}i}\tilde{A}^i}{s^2+\omega_\mathbf{k}^2},\;\;
\kappa_2=-8\Delta_0^3U\sum_{\mathbf{k}}\frac{u_\mathbf{k}}{\omega_\mathbf{k}}\frac{|f_\mathbf{k}|^2\mathcal{A}_{\mathbf{k}i}\tilde{A}^i}{s^2+\omega_\mathbf{k}^2}.
\end{align}

For sinusoidal external field $\tilde{\mathbf{A}}(s)\sim (s^2+\Omega^2)^{-1}$, which in frequency domain $s\rightarrow i\nu$ contributes the poles $\nu=\pm\Omega$. This describes the driving frequency. The resonance occurs when the poles from the driving frequency coincides with the poles from the collective modes. In frequency domain this comes from the denominators of \eqref{c11} to \eqref{kappa2}: $-\Omega^2+\omega_\mathbf{k}^2$. Recall that for the s wave pairing, when we focused on the Fermi surface, this factor becomes $-\Omega^2+4\Delta_0^2$ which can be pulled out from the summation or integration to give a simple pole at $\Omega=2\Delta_0$ (we consider only positive frequencies). For $d+id$, and in general beyond s wave, there is still $\mathbf{k}$ dependence even if we focused only on the Fermi surface. Specifically, the resonance from the collective modes comes from the factor $-\Omega^2+4\Delta_0^2|f_\mathbf{k}|^2$. The symmetry factor $f_\mathbf{k}\neq 1$ therefore broadens the resonance peak.

%

For the second order, the equations of motion give, after Laplace transformation
\begin{align}
\label{2ndordersyseq}
s\tilde{\sigma}^{x(2)}_\mathbf{k}+2\varepsilon_\mathbf{k}\tilde{\sigma}^{y(2)}_\mathbf{k}+2\Delta_0v_\mathbf{k}\tilde{\sigma}^{z(2)}_\mathbf{k}=&d_1\\
-2\varepsilon_\mathbf{k}\tilde{\sigma}^{x(2)}_\mathbf{k}+s\tilde{\sigma}^{y(2)}_\mathbf{k}-2\Delta_0u_\mathbf{k}\tilde{\sigma}^{z(2)}_\mathbf{k}=&d_2\\
-2\Delta_0v_\mathbf{k}\tilde{\sigma}^{x(2)}_\mathbf{k}+2\Delta_0u_\mathbf{k}\tilde{\sigma}^{y(2)}_\mathbf{k}+s\tilde{\sigma}^{z(2)}_\mathbf{k}=&d_3\\
\end{align}
where
\begin{align}
d_1=&-2L\{\Delta^{(1)}_I\sigma^{z(1)}_\mathbf{k}\}+4\Delta_0u_\mathbf{k}\mathcal{A}_{\mathbf{k}i}L\{A^i(t)\sigma^{z(1)}_\mathbf{k}\}+2\frac{\varepsilon_\mathbf{k}}{\omega_\mathbf{k}}\tilde{\Delta}^{(2)}_I-4\frac{\varepsilon_\mathbf{k}}{\omega_\mathbf{k}}\mathcal{A}_{\mathbf{k}i}L\{\Delta^{(1)}_RA^i(t)\}-\Delta_0\frac{v_\mathbf{k}}{\omega_\mathbf{k}}(\partial_i\partial_j\varepsilon_\mathbf{k})L\{A^i(t)A^j(t)\}\nonumber\\
&-4\Delta_0v_\mathbf{k}\frac{\varepsilon_\mathbf{k}}{\omega_\mathbf{k}}(g_{\mathbf{k},ij}+\mathcal{A}_{\mathbf{k}i}\mathcal{A}_{\mathbf{k}j})L\{A^i(t)A^j(t)\}\\
d_2=&2L\{\Delta^{(1)}_R\sigma^{z(1)}_\mathbf{k}\}+4\Delta_0v_\mathbf{k}\mathcal{A}_{\mathbf{k}i}L\{A^i(t)\sigma^{z(1)}_\mathbf{k}\}-4\frac{\varepsilon_\mathbf{k}}{\omega_\mathbf{k}}\mathcal{A}_{\mathbf{k}i}L\{\Delta^{(1)}_IA^i(t)\}-2\frac{\varepsilon_\mathbf{k}}{\omega_\mathbf{k}}\tilde{\Delta}^{(2)}_R+\Delta_0\frac{u_\mathbf{k}}{\omega_\mathbf{k}}(\partial_i\partial_j\varepsilon_\mathbf{k})L\{A^i(t)A^j(t)\}
\nonumber\\
&+4\Delta_0u_\mathbf{k}\frac{\varepsilon_\mathbf{k}}{\omega_\mathbf{k}}(g_{\mathbf{k},ij}+\mathcal{A}_{\mathbf{k}i}\mathcal{A}_{\mathbf{k}j})L\{A^i(t)A^j(t)\}\\
d_3=&-2L\{\Delta^{(1)}_R\sigma^{y(1)}_\mathbf{k}\}-4\Delta_0v_\mathbf{k}\mathcal{A}_{\mathbf{k}i}L\{A^i(t)\sigma^{y(1)}_\mathbf{k}\}-4\Delta_0\frac{v_\mathbf{k}}{\omega_\mathbf{k}}\mathcal{A}_{\mathbf{k}i}L\{\Delta^{(1)}_IA^i(t)\}+2L\{\Delta^{(1)}_I\sigma^{x(1)}_\mathbf{k}\}-2\Delta_0\frac{v_\mathbf{k}}{\omega_\mathbf{k}}\tilde{\Delta}^{(2)}_R\nonumber\\
&-4\Delta_0\frac{u_\mathbf{k}}{\omega_\mathbf{k}}\mathcal{A}_{\mathbf{k}i}L\{\Delta^{(1)}_RA^i(t)\}-4\Delta_0u_\mathbf{k}\mathcal{A}_{\mathbf{k}i}L\{A^i(t)\sigma^{x(1)}_\mathbf{k}\}+2\Delta_0\frac{u_\mathbf{k}}{\omega_\mathbf{k}}\tilde{\Delta}^{(2)}_I
\end{align}

We solve the system of equations \eqref{2ndordersyseq}. We only need the result for $\tilde{\sigma}^{x(2)}_\mathbf{k}$ and $\tilde{\sigma}^{y(2)}_\mathbf{k}$. We have
\begin{align}
\tilde{\sigma}^{x(2)}_\mathbf{k}=&\frac{1}{s(s^2+\omega^2_\mathbf{k})}\big[(s^2+4\Delta_0^2u^2_\mathbf{k})d_1+(4u_\mathbf{k}v_\mathbf{k}\Delta_0^2-2s\varepsilon_\mathbf{k})d_2-(2sv_\mathbf{k}\Delta_0+4u_\mathbf{k}\Delta_0\varepsilon_\mathbf{k})d_3\big]
\end{align}

\begin{align}
\tilde{\sigma}^{y(2)}_\mathbf{k}=&\frac{1}{s(s^2+\omega^2_\mathbf{k})}\big[(4u_\mathbf{k}v_\mathbf{k}\Delta_0^2+2s\varepsilon_\mathbf{k})d_1+(s^2+4\Delta_0^2v^2_\mathbf{k})d_2+(2su_\mathbf{k}\Delta_0-4v_\mathbf{k}\Delta_0\varepsilon_\mathbf{k})d_3\big].
\end{align}

The self-consistent equations for the Higgs mode is
\begin{align}
\tilde{\Delta}^{(2)}_R+i\tilde{\Delta}^{(2)}_I=-U\sum_\mathbf{k}(\tilde{\sigma}^{x(2)}_\mathbf{k}+i\tilde{\sigma}^{y(2)}_\mathbf{k}),
\end{align}
which gives, upon separating the real and imaginary parts, the system of equations
\begin{align}
[1-f_{R1}(s)]\tilde{\Delta}^{(2)}_R-f_{I1}(s)\tilde{\Delta}^{(2)}_I=&\mathcal{F}_1(s)\\
f_{R2}(s)\tilde{\Delta}^{(2)}_R+[1-f_{I2}(s)]\tilde{\Delta}^{(2)}_I=&\mathcal{F}_2(s).
\end{align}

We will define all the functions involve above shortly. The solution of the set of equations above is
\begin{align}
\tilde{\Delta}^{(2)}_R=&\frac{1}{\det\mathbb{M}_2}\bigg\{[1-f_{I2}(s)]\mathcal{F}_1(s)+f_{I1}(s)\mathcal{F}_2(s)\bigg\},
\end{align}
where
\begin{align}
\det\mathbb{M}_2=&[1-f_{R1}(s)][1-f_{I2}(s)]+f_{R2}(s)f_{I1}(s).
\end{align}

The necessary functions are given by
\begin{align}
f_{I1}(s)&=2U\sum_\mathbf{k}\frac{\varepsilon_\mathbf{k}s-2\Delta_0^2u_\mathbf{k}v_\mathbf{k}}{\omega_\mathbf{k}(s^2+\omega_\mathbf{k}^2)},\;\;
f_{R1}(s)=4U\sum_\mathbf{k}\frac{\Delta_0^2v_\mathbf{k}^2+\varepsilon_\mathbf{k}^2}{\omega_\mathbf{k}(s^2+\omega_\mathbf{k}^2)}\\
f_{I2}(s)&=4U\sum_\mathbf{k}\frac{\Delta_0^2u_\mathbf{k}^2+\varepsilon_\mathbf{k}^2}{\omega_\mathbf{k}(s^2+\omega_\mathbf{k}^2)},\;\;
f_{R2}(s)=2U\sum_\mathbf{k}\frac{2\Delta_0^2u_\mathbf{k}v_\mathbf{k}+\varepsilon_\mathbf{k}s}{\omega_\mathbf{k}(s^2+\omega_\mathbf{k}^2)}.
\end{align}

For the functions $\mathcal{F}_1(s)$ and $\mathcal{F}_2(s)$, we separate the three main contributions,
\begin{align}
\label{mathcalf}
\mathcal{F}_a(s)=F_a(s)_{con}+F_a(s)_{int}(s)+F_a(s)_{geo},
\end{align}
for $a=1,2$. We will give the explicit forms for $F_a(s)_{con}$, $F_a(s)_{int}$, and $F_a(s)_{geo}$ below, but first let us discuss the physical meaning of these terms.

The first function $F_a(s)_{con}$ is the conventional driving term and is proportional to the band curvature and second order driving $\propto L\{A^i(t)A^j(t)\}$. The 2nd term, $F_a(s)_{int}$, are the contributions of the products of the first order excitations. For example, it has contributions of the form $L\{A^i(t)\sigma^{j(1)}_\mathbf{k}\}$, which is a first order pseudospin excitation interacting with the external field. The explicit form will be shown below. The last term of \eqref{mathcalf} is the geometric driving term. It is proportional to the quantum metric and the product of Berry connection, $g_{\mathbf{k},ij}+\mathcal{A}_{\mathbf{k}i}\mathcal{A}_{\mathbf{k}j}$.

The explicit forms of these functions are
\begin{align}
F_1(s)_{con}=&-\Delta_0U\sum_\mathbf{k}\frac{v_\mathbf{k}s+2\varepsilon_\mathbf{k}u_\mathbf{k}}{\omega_\mathbf{k}(s^2+\omega_\mathbf{k}^2)}(\partial_i\partial_j\varepsilon_\mathbf{k})L\{A^i(t)A^j(t)\}\\
F_1(s)_{geo}=&-4\Delta_0U\sum_\mathbf{k}\frac{\varepsilon_\mathbf{k}(v_\mathbf{k}s+2\varepsilon_\mathbf{k}u_\mathbf{k})}{\omega_\mathbf{k}(s^2+\omega_\mathbf{k}^2)}(g_{\mathbf{k},ij}+\mathcal{A}_{\mathbf{k}i}\mathcal{A}_{\mathbf{k}j})L\{A^i(t)A^j(t)\}\\
F_1(s)_{int}=&U\sum_\mathbf{k}\frac{1}{s(s^2+\omega_\mathbf{k}^2)}\bigg[(s^2+4\Delta_0^2u_\mathbf{k}^2)\bigg(-2L\{\Delta^{(1)}_I\sigma^{z(1)}_\mathbf{k}\}+4\Delta_0u_\mathbf{k}\mathcal{A}_{\mathbf{k}i}L\{A^i(t)\sigma^{z(1)}_\mathbf{k}\}-4\frac{\varepsilon_\mathbf{k}}{\omega_\mathbf{k}}\mathcal{A}_{\mathbf{k}i}L\{\Delta^{(1)}_RA^i(t)\}\bigg)\nonumber\\
&+(4u_\mathbf{k}v_\mathbf{k}\Delta_0^2-2s\varepsilon_\mathbf{k})\bigg(2L\{\Delta^{(1)}_R\sigma^{z(1)}_\mathbf{k}\}+4\Delta_0v_\mathbf{k}\mathcal{A}_{\mathbf{k}i}L\{A^i(t)\sigma^{z(1)}_\mathbf{k}\}-4\frac{\varepsilon_\mathbf{k}}{\omega_\mathbf{k}}\mathcal{A}_{\mathbf{k}i}L\{\Delta^{(1)}_IA^i(t)\}\bigg)\nonumber\\
&-2\Delta_0(sv_\mathbf{k}+2u_\mathbf{k}\varepsilon_\mathbf{k})\bigg(-2L\{\Delta^{(1)}_R\sigma^{y(1)}_\mathbf{k}\}-4\Delta_0v_\mathbf{k}\mathcal{A}_{\mathbf{k}i}L\{A^i(t)\sigma^{y(1)}_\mathbf{k}\}-4\Delta_0\frac{v_\mathbf{k}}{\omega_\mathbf{k}}\mathcal{A}_{\mathbf{k}i}L\{\Delta^{(1)}_IA^i(t)\}\nonumber\\
&+2L\{\Delta^{(1)}_I\sigma^{x(1)}_\mathbf{k}\}-4\Delta_0\frac{u_\mathbf{k}}{\omega_\mathbf{k}}\mathcal{A}_{\mathbf{k}i}L\{\Delta^{(1)}_RA^i(t)\}-4\Delta_0u_\mathbf{k}\mathcal{A}_{\mathbf{k}i}L\{A^i(t)\sigma^{x(1)}_\mathbf{k}\}\bigg)\bigg]\\
F_2(s)_{con}=&\Delta_0U\sum_\mathbf{k}\frac{u_\mathbf{k}s-2\varepsilon_\mathbf{k}v_\mathbf{k}}{\omega_\mathbf{k}(s^2+\omega_\mathbf{k}^2)}(\partial_i\partial_j\varepsilon_\mathbf{k})L\{A^i(t)A^j(t)\}\\
F_2(s)_{geo}=&4\Delta_0U\sum_\mathbf{k}\frac{\varepsilon_\mathbf{k}(u_\mathbf{k}s-2\varepsilon_\mathbf{k}v_\mathbf{k})}{\omega_\mathbf{k}(s^2+\omega_\mathbf{k}^2)}(g_{\mathbf{k},ij}+\mathcal{A}_{\mathbf{k}i}\mathcal{A}_{\mathbf{k}j})L\{A^i(t)A^j(t)\}
\end{align}

\begin{align}
F_2(s)_{int}=&U\sum_\mathbf{k}\frac{1}{s(s^2+\omega_\mathbf{k}^2)}\bigg[(4u_\mathbf{k}v_\mathbf{k}\Delta_0^2+2s\varepsilon_\mathbf{k})\bigg(-2L\{\Delta^{(1)}_I\sigma^{z(1)}_\mathbf{k}+4\Delta_0u_\mathbf{k}\mathcal{A}_{\mathbf{k}i}L\{A^i(t)\sigma^{z(1)}_\mathbf{k}\}-4\frac{\varepsilon_\mathbf{k}}{\omega_\mathbf{k}}\mathcal{A}_{\mathbf{k}i}L\{\Delta^{(1)}_RA^i(t)\}\bigg)\nonumber\\
&+(s^2+4v_\mathbf{k}^2\Delta_0^2)\bigg(2L\{\Delta^{(1)}_R\sigma^{z(1)}_\mathbf{k}\}+4\Delta_0v_\mathbf{k}\mathcal{A}_{\mathbf{k}i}L\{A^i(t)\sigma^{z(1)}_\mathbf{k}\}-4\frac{\varepsilon_\mathbf{k}}{\omega_\mathbf{k}}\mathcal{A}_{\mathbf{k}i}L\{\Delta^{(1)}_IA^i(t)\}\bigg)\nonumber\\
&+2\Delta_0(su_\mathbf{k}-2v_\mathbf{k}\varepsilon_\mathbf{k})\bigg(-2L\{\Delta^{(1)}_R\sigma^{y(1)}_\mathbf{k}\}-4\Delta_0v_\mathbf{k}\mathcal{A}_{\mathbf{k}i}L\{A^i(t)\sigma^{y(1)}_\mathbf{k}\}-4\Delta_0\frac{v_\mathbf{k}}{\omega_\mathbf{k}}\mathcal{A}_{\mathbf{k}i}L\{\Delta^{(1)}_IA^i(t)\}\nonumber\\
&+2L\{\Delta^{(1)}_I\sigma^{x(1)}_\mathbf{k}\}-4\Delta_0\frac{u_\mathbf{k}}{\omega_\mathbf{k}}\mathcal{A}_{\mathbf{k}i}L\{\Delta^{(1)}_RA^i(t)\}-4\Delta_0u_\mathbf{k}\mathcal{A}_{\mathbf{k}i}L\{A^i(t)\sigma^{x(1)}_\mathbf{k}\}\bigg).
\end{align}


The Laplace transforms for the products of first order contributions are
\begin{align}
L\{\Delta^{(1)}_C\sigma^{z(1)}_\mathbf{k}\}
&=\sum_{m=1}^2\sum_{n=1}^4\frac{d_{Cm}s_{zn}\delta_{Cm}(\delta_{Cm}^2-\omega_n^2+s^2)}{\delta_{Cm}^4+2\delta_{Cm}^2(s^2-\omega_n^2)+(s^2+\omega_n^2)^2}\\
L\{A^i(t)\sigma^{z(1)}_\mathbf{k}\}&=\sum_{n=1}^4\frac{A^is_{zn}\Omega(\Omega^2-\omega_n^2+s^2)}{\Omega^4+2\Omega^2(s^2-\omega_n^2)+(s^2+\omega_n^2)^2}\\
L\{A^i(t)\sigma^{y(1)}_\mathbf{k}\}&=\sum_{n=1}^4\frac{2A^is_{yn}\Omega\omega_ns}{\Omega^4+2\Omega^2(s^2-\omega_n^2)+(s^2+\omega_n^2)^2}\\
L\{\Delta^{(1)}_CA^i(t)\}&=\sum_{n=1}^4\frac{2A^id_{Cn}\Omega\delta_ns}{\Omega^4+2\Omega^2(s^2-\delta_n^2)+(s^2+\delta_n^2)^2},
\end{align}
where $C=R,I$ denotes real and imaginary parts, respectively. The other variables are given by $\delta_{I1}=\Omega_2$, $\delta_{I2}=\Omega, \delta_{R1}=\Omega_1$, $\delta_{R2}=\Omega$, and $\omega_{1,2,3,4}=\bar{\omega}$, $\Omega_1$, $\Omega_2$, and $\Omega$. Here $\Omega_1$ is given by
\begin{align}
\label{polefirst}
\Omega_1=2\Delta_0\sqrt{\overline{|f|}^2-\frac{Uh_2}{8\Delta_0}},
\end{align}
while $\Omega_2^2=\bar{\omega}^2-2\Delta_0Uh_4$. The functions $h_{1i}$, $h_{2}$, $h_{3i}$, and $h_{4}$ are to be defined below in \eqref{hs}.

The different factors appearing above are as follows. For $s_{yn}$:
\begin{align}
s_{y1}&=-\nu_\mathbf{k}\frac{u_\mathbf{k}}{\omega_\mathbf{k}}\frac{16\Delta_0^4Uh_{1i}A^i\Omega}{\bar{\omega}(\bar{\omega}^2-\Omega_1^2)(\bar{\omega}^2-\Omega^2)}-\frac{u_\mathbf{k}^2}{\omega_\mathbf{k}}\frac{32\Delta_0^5h_{3i}A^i\Omega}{\bar{\omega}(\bar{\omega}^2-\Omega_2^2)(\bar{\omega}^2-\Omega^2)}+\frac{u_\mathbf{k}}{\omega_\mathbf{k}}\frac{8\Delta_0^3|f_\mathbf{k}|^2\Omega}{\bar{\omega}(\bar{\omega}^2-\Omega^2)}\mathcal{A}_{\mathbf{k}i}A^i\\
s_{y2}&=\nu_\mathbf{k}\frac{u_\mathbf{k}}{\omega_\mathbf{k}}\frac{16\Delta_0^4Uh_{1i}A^i\Omega}{\Omega_1(\bar{\omega}^2-\Omega_1^2)(\Omega_1^2-\Omega^2)}\\
s_{y3}&=\frac{u_\mathbf{k}^2}{\omega_\mathbf{k}}\frac{32\Delta_0^5h_{3i}A^i\Omega}{\Omega_2(\bar{\omega}^2-\Omega_2^2)(\Omega_2^2-\Omega^2)}\\
s_{y4}&=\nu_\mathbf{k}\frac{u_\mathbf{k}}{\omega_\mathbf{k}}\frac{16\Delta_0^4Uh_{1i}A^i}{(\bar{\omega}^2-\Omega^2)(-\Omega_1^2+\Omega^2)}+\frac{u_\mathbf{k}^2}{\omega_\mathbf{k}}\frac{32\Delta_0^5h_{3i}A^i}{(\bar{\omega}^2-\Omega^2)(-\Omega_2^2+\Omega^2)}-\frac{u_\mathbf{k}}{\omega_\mathbf{k}}\frac{8\Delta_0^3|f_\mathbf{k}|^2}{\bar{\omega}^2-\Omega^2}\mathcal{A}_{\mathbf{k}i}A^i.
\end{align}

For $s_{zn}$:
\begin{align}
s_{z1}&=-\frac{\nu_\mathbf{k}}{\omega_\mathbf{k}}\frac{8\Delta_0^3Uh_{1i}A^i\Omega}{(\bar{\omega}^2-\Omega_1^2)(\bar{\omega}^2-\Omega^2)}-\frac{u_\mathbf{k}}{\omega_\mathbf{k}}\frac{16\Delta_0^4h_{3i}A^i}{(\bar{\omega}^2-\Omega_2^2)(\bar{\omega}^2-\Omega^2)}-\frac{|f_\mathbf{k}|^2}{\omega_\mathbf{k}}\frac{4\Delta_0^2\mathcal{A}_{\mathbf{k}i}A^i}{\Omega^2-\bar{\omega}^2}\\
s_{z2}&=-\frac{\nu_\mathbf{k}}{\omega_\mathbf{k}}\frac{8\Delta_0^3Uh_{1i}A^i\Omega}{(\Omega_1^2-\bar{\omega}^2)(\Omega_1^2-\Omega^2)}\\
s_{z3}&=-\frac{u_\mathbf{k}}{\omega_\mathbf{k}}\frac{16\Delta_0^4h_{3i}A^i}{(\Omega_2^2-\bar{\omega}^2)(\Omega_2^2-\Omega^2)}\\
s_{z4}&=-\frac{\nu_\mathbf{k}}{\omega_\mathbf{k}}\frac{8\Delta_0^3Uh_{1i}A^i\Omega}{(\Omega^2-\bar{\omega}^2)(\Omega^2-\Omega_1^2)}-\frac{u_\mathbf{k}}{\omega_\mathbf{k}}\frac{16\Delta_0^4h_{3i}A^i}{(\Omega^2-\bar{\omega}^2)(\Omega^2-\Omega_2^2)}-\frac{|f_\mathbf{k}|^2}{\omega_\mathbf{k}}\frac{4\Delta_0^2\mathcal{A}_{\mathbf{k}i}A^i}{\bar{\omega}^2-\Omega^2}.
\end{align}

The $d_{Cm}$s:
\begin{align}
d_{R1}=\frac{4\Delta_0^2Uh_{1i}A^i\Omega}{\Omega_1(\Omega^2-\Omega_1^2)},\;\;d_{R2}=\frac{4\Delta_0^2Uh_{1i}A^i}{\Omega_1^2-\Omega^2},\;\;
d_{I1}=-\frac{8\Delta_0^3Uh_{3i}A^i}{\Omega_2(\Omega^2-\Omega_2^2)},\;\;d_{I2}=-\frac{8\Delta_0^3Uh_{3i}A^i}{\Omega(\Omega_2^2-\Omega^2)}.
\end{align}

The $h$s are defined as
\begin{align}
\label{hs}
h_{i1}=\sum_\mathbf{k}\nu_\mathbf{k}|f_\mathbf{k}|\mathcal{A}_{\mathbf{k}i},\;\;h_2=\sum_\mathbf{k}\frac{\nu_\mathbf{k}^2}{|f_\mathbf{k}|},\;\;
h_{3i}=\sum_\mathbf{k}\frac{u_\mathbf{k}}{\omega_\mathbf{k}}|f_\mathbf{k}|^2\mathcal{A}_{\mathbf{k}i},\;\;
h_4=\sum_\mathbf{k}\frac{u_\mathbf{k}^2}{|f_\mathbf{k}|}.
\end{align}

We perform analytic continuation to go from Laplace space to frequency space $s\rightarrow i\nu$. This gives the different poles
\begin{align}
\nu_{n,Im}=\omega_n\pm\delta_{Im},\;
\nu_n=\omega_n\pm\Omega,\;
\nu_{Im}=\Omega\pm\delta_{Im}.
\end{align}

For s wave ($u_\mathbf{k}=1$ and $v_\mathbf{k}=0$), the Anderson resonance comes from the factor $s^2+\omega_\mathbf{k}^2$ after analytic continuation $s\rightarrow 2i\Omega$ (the driving resonance is $\nu=2\Omega$ for 2nd order).

Of particular importance is the poles due to the coupling of $\Delta_I^{(1)}(t)$ and the external field $\mathbf{A}(t)$ to give a first order Higgs mode discussed in the main text. This gives coincident poles at $\frac{1}{2}(\bar{\omega}+\Omega_2)$, $\frac{1}{2}(\Omega_1+\Omega_2)$, and $\Omega_2$, This effectively gives a 3rd order pole which results in a very strong peak. 

\section{Gauge invariance.} 
\label{gaugeinvariance}
In this section we discuss the issue of gauge invariance of our theory. The important point to keep in mind here is that in the second quantization formalism we have to perform gauge transformations both on the operator and its associated wavefunction \cite{Greiter2005}. For example, if we expand an $N$-electron state $|\phi\rangle$ in terms of the field operators $\{\psi^\dagger_{\sigma_i}(\mathbf{x}_1)\}$, we have
\begin{align}
|\phi\rangle =&\sum_{\sigma_1\cdot\cdot\cdot\sigma_N}\int d^3\mathbf{x}_1\cdot\cdot\cdot d^3\mathbf{x}_N\phi(\mathbf{x}_1\cdot\cdot\cdot\mathbf{x}_N;\sigma_1\cdot\cdot\cdot\sigma_N)\psi^\dagger_{\sigma_1}(\mathbf{x}_1)\cdot\cdot\cdot\psi^\dagger_{\sigma_N}(\mathbf{x}_N)|0\rangle.
\end{align}

This state is invariant under the gauge transformation of the field operators
\begin{eqnarray}
\psi^\dagger_\sigma(\mathbf{x})\rightarrow e^{i\theta(\mathbf{x})}\psi^\dagger_\sigma(\mathbf{x})
\end{eqnarray}
if we also simultaneously transform the many-electron wavefunction
\begin{align}
\label{wfgauge}
\phi(\mathbf{x}_1\cdot\cdot\cdot\mathbf{x}_N;\sigma_1\cdot\cdot\cdot\sigma_N)\rightarrow\prod_{j=1}^Ne^{-i\theta(\mathbf{x}_j)}\phi(\mathbf{x}_1\cdot\cdot\cdot\mathbf{x}_N;\sigma_1\cdot\cdot\cdot\sigma_N).
\end{align}

Another example is the BCS groundstate
\begin{eqnarray}
|\psi_\phi\rangle=\prod_\mathbf{k}(u_\mathbf{k}+v_\mathbf{k}e^{i\phi}c^\dagger_{\mathbf{k}\uparrow}c^\dagger_{-\mathbf{k}\downarrow})|0\rangle .
\end{eqnarray}
This is gauge invariant under the set of gauge transformations
\begin{align}
\label{phigauge}
c^\dagger_{\mathbf{k}\uparrow}\rightarrow e^{i\theta_{\mathbf{k}\uparrow}}c^\dagger_{\mathbf{k}\uparrow},\;\;\;\;
c^\dagger_{-\mathbf{k}\downarrow}\rightarrow e^{i\theta_{-\mathbf{k}\downarrow}}c^\dagger_{-\mathbf{k}\downarrow},\;\;\;\;
\phi \rightarrow \phi-\theta_{\mathbf{k}\uparrow}-\theta_{-\mathbf{k}\downarrow}.
\end{align}

Note that $\phi$, which can depend on $\mathbf{k}$, is the phase of the Cooper pair wavefunction. Hence, \eqref{phigauge} actually comes from the gauge transformation of the two-electron wavefunction, which is a special case of \eqref{wfgauge}. In the main text, we fixed the gauge by choosing this phase to be zero.

The interaction Hamiltonian responsible for superconductivity has the general form
\begin{eqnarray}
\label{hi}
H_I&=&\frac{1}{2}\sum_{lmij}V_{lmij}b^\dagger_jb^\dagger_mb_lb_i
\end{eqnarray}
where
\begin{align}
\label{vmatrix}
V_{lmij}=
\int d^3\mathbf{r}_1&d^3\mathbf{r}_2\phi^*_m(\mathbf{r}_1)\phi_l(\mathbf{r}_1)V(\mathbf{r}_1-\mathbf{r}_2)\phi^*_j(\mathbf{r}_2)\phi_i(\mathbf{r}_2)
\end{align}
with the indices $\{lmij\}$ denoting the collective indices for the band, wavevector, and spin.

The Hamiltonian is invariant under simultaneous gauge transformation $b_i\rightarrow e^{-i\theta_i}b_i$ and $\phi_i\rightarrow e^{i\theta_i}\phi_i$ as it should be.

Let us now apply this prescription to the reduced BCS Hamiltonian pairing term
\begin{eqnarray}
H_\Delta =-\sum_\mathbf{k}(\Delta_\mathbf{k}c_{\mathbf{k}\uparrow}^\dagger c_{\mathbf{k}\uparrow}^\dagger +h.c.)
\end{eqnarray}
where the pairing potential satisfies the self-consistency condition
\begin{eqnarray}
\Delta_\mathbf{k}=-\sum_\mathbf{p}V_{\mathbf{k}\mathbf{p}}\langle c_{-\mathbf{p}\downarrow} c_{\mathbf{p}\uparrow}\rangle.
\end{eqnarray}

Note that $V_{\mathbf{k}\mathbf{p}}$ is a special case of $V_{lmij}$ in \eqref{vmatrix}. Eq. \eqref{hi} is invariant under the set of gauge transformations
\begin{align}
c_{\mathbf{p}\uparrow}&\rightarrow e^{-i\theta_{\mathbf{p}\uparrow}}c_{\mathbf{p}\uparrow},\;\; c_{-\mathbf{p}\downarrow}\rightarrow e^{-i\theta_{-\mathbf{p}\downarrow}}c_{-\mathbf{p}\downarrow},\;\;
c^\dagger_{\mathbf{p}\uparrow}\rightarrow e^{i\theta_{\mathbf{p}\uparrow}}c^\dagger_{\mathbf{p}\uparrow},\nonumber\\ c^\dagger_{-\mathbf{p}\downarrow}&\rightarrow e^{i\theta_{-\mathbf{p}\downarrow}}c^\dagger_{-\mathbf{p}\downarrow},\;\;
V_{\mathbf{k}\mathbf{p}}\rightarrow e^{i\theta_{\mathbf{p}\uparrow}}e^{i\theta_{-\mathbf{p}\downarrow}}e^{-i\theta_{\mathbf{k}\uparrow}}e^{-i\theta_{-\mathbf{k}\downarrow}}V_{\mathbf{k}\mathbf{p}}.\nonumber
\end{align}

\textit{Gauge transformation of pseudospins.} Recall the definition of speudospin
\begin{eqnarray}
\vec{\sigma}= \frac{1}{2}\psi^\dagger_\mathbf{k}\vec{\tau}\psi_\mathbf{k}.
\end{eqnarray}

Let us see how this change under a gauge transformation. The x component becomes
\begin{align}
\sigma^x_\mathbf{k}=\frac{1}{2}(c^\dagger_{\mathbf{k}\uparrow}c^\dagger_{-\mathbf{k}\downarrow}+c_{-\mathbf{k}\downarrow}c_{\mathbf{k}\uparrow})
\rightarrow &\frac{1}{2}(c^\dagger_{\mathbf{k}\uparrow}e^{-i(\theta_{\mathbf{k}\uparrow}+\theta_{-\mathbf{k}\downarrow})}c^\dagger_{-\mathbf{k}\downarrow}+c_{-\mathbf{k}\downarrow}e^{i(\theta_{\mathbf{k}\uparrow}+\theta_{-\mathbf{k}\downarrow})}c_{\mathbf{k}\uparrow})\nonumber\\
=&\cos(\theta_{\mathbf{k}\uparrow}+\theta_{-\mathbf{k}\downarrow})\sigma^x_\mathbf{k}+\sin(\theta_{\mathbf{k}\uparrow}+\theta_{-\mathbf{k}\downarrow})\sigma^y_\mathbf{k}.
\end{align}
Similar calculations for the y component yields
\begin{eqnarray}
\sigma^y_\mathbf{k}\rightarrow -\sin(\theta_{\mathbf{k}\uparrow}+\theta_{-\mathbf{k}\downarrow})\sigma^x_\mathbf{k}+\cos(\theta_{\mathbf{k}\uparrow}+\theta_{-\mathbf{k}\downarrow})\sigma^y_\mathbf{k}.
\end{eqnarray}
Hence, the gauge transformation acts as a rotation about the z axis for the pseudospins
\begin{eqnarray}
\begin{pmatrix}
\sigma^x_\mathbf{k}\\
\sigma^y_\mathbf{k}
\end{pmatrix}
\rightarrow
\begin{pmatrix}
\cos\alpha_\mathbf{k} & \sin\alpha_\mathbf{k}\\
-\sin\alpha_\mathbf{k} & \cos\alpha_\mathbf{k}
\end{pmatrix}
\begin{pmatrix}
\sigma^x_\mathbf{k}\\
\sigma^y_\mathbf{k}
\end{pmatrix}
\end{eqnarray}
with the rotation angle given by $\alpha_\mathbf{k}=\theta_{\mathbf{k}\uparrow}+\theta_{-\mathbf{k}\downarrow}$.

This shows that the pseudospin formalism is \textit{not} manifestly gauge invariant as the pseudospin variables transform non-trivially under such transformation. However, as we show below, the Higgs mode is gauge invariant.

\textit{Multiband superconductor.} Let us consider the hopping part of a tight-binding Hamiltonian ignoring first the Peierls substitution. The hopping amplitude matrix elements are given by
\begin{eqnarray}
K^\sigma_{i\alpha ,j\beta}=\int d\mathbf{r}\phi^*_{i\alpha\sigma}(\mathbf{r})\hat{K}\phi_{j\beta\sigma}(\mathbf{r}).
\end{eqnarray}

The hopping Hamiltonian is invariant under the gauge transformation
\begin{align}
c_{j\beta\sigma}\rightarrow e^{-i\theta_{j\beta\sigma}}c_{j\beta\sigma},\;\;
c^{\dagger}_{i\alpha\sigma}\rightarrow e^{i\theta_{i\alpha\sigma}}c^\dagger_{i\alpha\sigma}
\phi_{j\beta\sigma}\rightarrow e^{i\theta_{j\beta\sigma}}\phi_{j\beta\sigma},\;\;
\phi^*_{i\alpha\sigma}\rightarrow e^{-i\theta_{i\alpha\sigma}}\phi^*_{i\alpha\sigma}.
\end{align}

Note that under this transformation the hopping amplitude transforms as
\begin{eqnarray}
K^\sigma_{i\alpha ,j\beta}\rightarrow e^{-i\theta_{i\alpha\sigma}}K^\sigma_{i\alpha ,j\beta}e^{i\theta_{j\beta\sigma}}.
\end{eqnarray}
This is consistent with the gauge transformation of a Wilson link of a lattice gauge theory.

Similarly in band space
\begin{eqnarray}
H_K=\sum_\mathbf{k}d^{\dagger \;n}_\mathbf{k}\tilde{K}(\mathbf{k})_n^{\;m}d_{\mathbf{k}m}
=\sum_\mathbf{k}d^{\dagger \;n}_\mathbf{k}\mathcal{G}^{\dagger\;\alpha}_{\mathbf{k}n}\tilde{K}(\mathbf{k})_\alpha^{\;\beta}\mathcal{G}^{\;\;m}_{\mathbf{k}\beta}d_{\mathbf{k}m}.
\end{eqnarray}

The gauge transformation for the matrix
\begin{eqnarray}
\tilde{K}(\mathbf{k})_n^{\;m}\rightarrow e^{-i\theta_{\mathbf{k}n}}\tilde{K}(\mathbf{k})_n^{\;m}e^{i\theta_{\mathbf{k}m}}
\end{eqnarray}
can be thought-of as coming from the gauge transformation of the Bloch functions
\begin{eqnarray}
\mathcal{G}^{\;\;m}_{\mathbf{k}\beta}\rightarrow e^{i\theta_{\mathbf{k}m}}\mathcal{G}^{\;\;m}_{\mathbf{k}\beta}\;\;\;\mbox{and}\;\;\;
\label{gtransform}
\mathcal{G}^{\dagger\;\alpha}_{\mathbf{k}n}\rightarrow e^{-i\theta_{\mathbf{k}n}}\mathcal{G}^{\dagger\;\alpha}_{\mathbf{k}n},
\end{eqnarray}
for fixed $m$ and $n$.

Note that above, we are careful in using lowered indices to label rows and raised indices to label columns. This will be useful later. The pairing Hamiltonian, written in band space has a term of the form
\begin{eqnarray}
d^{\;n}_{-\mathbf{k}\downarrow}\Delta_{\mathbf{k}n}^{\;\;m}d_{\mathbf{k}\uparrow m}=d^{\;n}_{-\mathbf{k}\downarrow}\mathcal{G}^{\dagger\;\alpha}_{\mathbf{k}n}\Delta_\alpha^{\;\beta}\mathcal{G}^{\;\;m}_{\mathbf{k}\beta}d_{\mathbf{k}\uparrow m}.
\end{eqnarray}

This invariant under 
\begin{eqnarray}
d_{\mathbf{k}\uparrow m}&\rightarrow e^{-i\theta_{\mathbf{k}m}}d_{\mathbf{k}\uparrow m},\;\;d^{\;n}_{-\mathbf{k}\downarrow}&\rightarrow e^{i\theta_{\mathbf{k}n}}d^{\;n}_{-\mathbf{k}\downarrow}
\end{eqnarray}
along with \eqref{gtransform}.

\textit{In the presence of vector potential.} Let us now consider the gauge transformation in real space in the presence of a vector potential. The hopping matrix with Peierls substitution transforms as
\begin{align}
K^\sigma_{i\alpha ,j\beta}e^{i\mathbf{A}\cdot(\mathbf{r}_{i\alpha}-\mathbf{r}_{j\beta})}\rightarrow e^{-i\theta_{i\alpha}}K^\sigma_{i\alpha ,j\beta}e^{i\mathbf{A}\cdot(\mathbf{r}_{i\alpha}-\mathbf{r}_{j\beta})}e^{i\theta_{j\beta}}
=K^\sigma_{i\alpha ,j\beta}\exp\left\{ i\left[\mathbf{A}-\frac{(\theta_{i\alpha}-\theta_{j\beta})}{R_{i\alpha,j\beta}}\hat{R}_{i\alpha,j\beta}\right]\cdot\mathbf{R}_{i\alpha,j\beta}\right\}
\end{align}
where $\mathbf{R}_{i\alpha,j\beta}=\mathbf{r}_{i\alpha}-\mathbf{r}_{j\beta}$, with $\hat{R}_{i\alpha,j\beta}$ and $R_{i\alpha,j\beta}$ its corresponding unit vector and magnitude.

In the continuum limit $R_{i\alpha,j\beta}\rightarrow 0$ we get the familiar gauge transformation
\begin{eqnarray}
\mathbf{A}-\frac{(\theta_{i\alpha}-\theta_{j\beta})}{R_{i\alpha,j\beta}}\hat{R}_{i\alpha,j\beta}\longrightarrow \mathbf{A}-\nabla\theta.
\end{eqnarray}

The tight-binding Hamiltonian in Fourier space has the form
\begin{align}
H_\mathbf{k}(\mathbf{A})
&=\sum_{\mathbf{k}\alpha\beta}c^{\dagger\;\alpha}_{\mathbf{k}\sigma}e^{-i(\mathbf{k}-\mathbf{A})\cdot\delta_\alpha}\tilde{K}^\sigma(\mathbf{k}-\mathbf{A})_\alpha^{\;\beta}e^{i(\mathbf{k}-\mathbf{A})\cdot\delta_\beta}c_{\mathbf{k}\beta\sigma}.
\end{align}

Note that the exponential factors are just gauge transformation, i.e. they can be gauged away. The Hamiltonian can therefore be written as
\begin{align}
H_\mathbf{k}(\mathbf{A})=\sum_{\mathbf{k}\alpha\beta}c^{\dagger\alpha}_{\mathbf{k}\sigma}\tilde{K}^\sigma(\mathbf{k}-\mathbf{A})_\alpha^{\;\beta}c_{\mathbf{k}\beta\sigma}=\sum_{\mathbf{k}nm}d^{\dagger n}_{\mathbf{k}\sigma}\tilde{K}^\sigma(\mathbf{k}-\mathbf{A})_n^{\;m}d_{\mathbf{k}m\sigma}
\end{align}
in orbital and band spaces, respectively.

This is gauge invariant under
\begin{align}
d_{\mathbf{k}m\sigma}\rightarrow e^{-i\theta_{\mathbf{k}-\mathbf{A}m\sigma}}d_{\mathbf{k}m\sigma},\;\;
d^{\dagger\;n}_{\mathbf{k}\sigma}\rightarrow e^{i\theta_{\mathbf{k}-\mathbf{A}n\sigma}}d^{\dagger\;n}_{\mathbf{k}\sigma},\;\;
\tilde{K}^\sigma(\mathbf{k}-\mathbf{A})_n^{\;m}\rightarrow e^{-i\theta_{\mathbf{k}-\mathbf{A}n\sigma}}\tilde{K}^\sigma(\mathbf{k}-\mathbf{A})_n^{\;m}e^{i\theta_{\mathbf{k}-\mathbf{A}m\sigma}}\nonumber.
\end{align}

The Nambu spinor is written as
\begin{eqnarray}
\psi_{\mathbf{k},n}=
\begin{pmatrix}
d_{\mathbf{k}\uparrow n}\\
d^\dagger_{-\mathbf{k}\downarrow n}
\end{pmatrix}=
\begin{pmatrix}
\mathcal{G}^{\dagger\;\alpha}_{\mathbf{k}-\mathbf{A}n}c_{\mathbf{k}\uparrow\alpha}\\
\mathcal{G}^{\dagger\;\alpha}_{\mathbf{k}+\mathbf{A}n}c_{-\mathbf{k}\downarrow\alpha}
\end{pmatrix}.
\end{eqnarray}

Note that since this is written as a column vector, the creation operators are written with lowered band and orbital indices as they label rows. Another way to think of the bottom-half row is that they are annihilation operators for the holes.

The pairing Hamiltonian contains the term of the form
\begin{eqnarray}
d_{-\mathbf{k}\downarrow}^{\;n}\Delta_n^{\;m}d_{\mathbf{k}\uparrow m}=d_{-\mathbf{k}\downarrow}^{\;n}\mathcal{G}^{\dagger\;\alpha}_{\mathbf{k}+\mathbf{A}n}\Delta_\alpha^{\;\beta}\mathcal{G}^{\;m}_{\mathbf{k}-\mathbf{A}\beta}d_{\mathbf{k}\uparrow m}.
\end{eqnarray}

This is invariant under
\begin{align}
d_{\mathbf{k}\uparrow m}\rightarrow e^{-i\theta_{\mathbf{k}-\mathbf{A}m\uparrow}}d_{\mathbf{k}\uparrow m},\;
d^{\;n}_{-\mathbf{k}\downarrow}\rightarrow e^{i\theta_{\mathbf{k}+\mathbf{A}n\downarrow}}d^{\;n}_{-\mathbf{k}\downarrow},\;
\mathcal{G}^{\;m}_{\mathbf{k}-\mathbf{A}\beta}\rightarrow e^{i\theta_{\mathbf{k}-\mathbf{A}m\uparrow}}\mathcal{G}^{\;m}_{\mathbf{k}-\mathbf{A}\beta},\;
\mathcal{G}^{\dagger\;\alpha}_{\mathbf{k}+\mathbf{A}n}\rightarrow e^{-i\theta_{\mathbf{k}+\mathbf{A}n\downarrow}}\mathcal{G}^{\dagger\;\alpha}_{\mathbf{k}+\mathbf{A}n}.\nonumber
\end{align}

In the pseudospin formalism, we expand in powers of the vector potential. Let us see how this gauge transformation looks like when this is done on a Bloch function. That is, $\mathcal{G}$ with fixed band index $m$, or a column element of $\mathcal{G}$. We denote this by $\bar{\mathcal{G}}$. We have
\begin{eqnarray}
\bar{\mathcal{G}}_{\mathbf{k}-\mathbf{A}}e^{i\theta_{\mathbf{k}-\mathbf{A}}}&=&(\bar{\mathcal{G}}_\mathbf{k}-A^a(\partial_a\bar{\mathcal{G}}_\mathbf{k})-iA^a(\partial_a\theta_\mathbf{k})\bar{\mathcal{G}}_\mathbf{k}+\cdot\cdot\cdot)e^{i\theta_\mathbf{k}}\nonumber\\
\bar{\mathcal{G}}^\dagger_{\mathbf{k}+\mathbf{A}}e^{-i\theta_{\mathbf{k}+\mathbf{A}}}&=&(\bar{\mathcal{G}}^\dagger_\mathbf{k}+A^a(\partial_a\bar{\mathcal{G}}^\dagger_\mathbf{k})-iA^a(\partial_a\theta_\mathbf{k})\bar{\mathcal{G}}^\dagger_\mathbf{k}+\cdot\cdot\cdot)e^{-i\theta_\mathbf{k}}.\nonumber
\end{eqnarray}

We are mostly interested in the case $\Delta_\alpha^{\;\beta}=\Delta\delta_\alpha^{\;\beta}$, for which we have
\begin{align}
\bar{\mathcal{G}}^\dagger_{\mathbf{k}+\mathbf{A}}e^{-i\theta_{\mathbf{k}+\mathbf{A}}}\Delta\bar{\mathcal{G}}_{\mathbf{k}-\mathbf{A}}e^{i\theta_{\mathbf{k}-\mathbf{A}}}=\Delta[1+2iA^a(\mathcal{A}_{\mathbf{k}a}-\partial_a\theta_\mathbf{k})+\cdot\cdot\cdot]
\end{align}
where $\vec{\mathcal{A}}_\mathbf{k}$ is the Berry connection. Hence, we see the known gauge transformation of the Berry connection $\vec{\mathcal{A}}_\mathbf{k}-\nabla_\mathbf{k}\theta_\mathbf{k}$.

Let us focus on a single band with Bloch function given by the column vector $\mathcal{G}_{\mathbf{k}\beta}$. Here we suppressed the band index as we are considering only a single band. 

The order parameter in band space is given by and transforms as
\begin{align}
\Delta_\mathbf{k}(\mathbf{A})=\mathcal{G}^\dagger_{\mathbf{k}+\mathbf{A}}\Delta_\alpha^{\;\beta}\mathcal{G}_{\mathbf{k}-\mathbf{A}\beta}\rightarrow e^{-i\theta_{\mathbf{k}+\mathbf{A}\downarrow}}\mathcal{G}^\dagger_{\mathbf{k}+\mathbf{A}}\Delta_\alpha^{\;\beta}\mathcal{G}_{\mathbf{k}-\mathbf{A}\beta}e^{i\theta_{\mathbf{k}-\mathbf{A}\uparrow}}.\nonumber
\end{align}
It only acquires a phase which makes it clear that the magnitude is gauge invariant. However, it is instructive to show the gauge invariance of the magnitude order-by-order in the expansion in powers of $\mathbf{A}$ since our method of calculating the Higgs mode uses this expansion.

Before the gauge transformation, we have the following expansion
\begin{align}
\label{deltaexpand1}
\mathcal{G}^\dagger_{\mathbf{k}+\mathbf{A}}\Delta_\alpha^{\;\beta}\mathcal{G}_{\mathbf{k}-\mathbf{A}\beta}=\Delta +2i\Delta\mathcal{A}_{\mathbf{k}j}A^j-2\Delta(g_{\mathbf{k},ij}+\mathcal{A}_{\mathbf{k}i}\mathcal{A}_{\mathbf{k}j})A^iA^j
\end{align}
where $g_{\mathbf{k},ij}$ is the quantum metric.

In expanding the gauged-transformed order parameter, we need the following expansions
\begin{align}
\label{exp1}
e^{i\theta_{\mathbf{k}-\mathbf{A}\uparrow}}&=e^{i\theta_{\mathbf{k}\uparrow}}\bigg[1-iA^j\partial_j\theta_{\mathbf{k}\uparrow}-\frac{1}{2}(\partial_j\theta_{\mathbf{k}\uparrow})(\partial_l\theta_{\mathbf{k}\uparrow})A^jA^l+\frac{i}{2}A^jA^l\partial_j\partial_l\theta_{\mathbf{k}\uparrow}+\cdot\cdot\cdot\bigg]\\
\label{exp2}
e^{-i\theta_{\mathbf{k}+\mathbf{A}\downarrow}}&=e^{-i\theta_{\mathbf{k}\downarrow}}\bigg[1-iA^j\partial_j\theta_{\mathbf{k}\downarrow}-\frac{1}{2}(\partial_j\theta_{\mathbf{k}\downarrow})(\partial_l\theta_{\mathbf{k}\downarrow})A^jA^l-\frac{i}{2}A^jA^l\partial_j\partial_l\theta_{\mathbf{k}\downarrow}+\cdot\cdot\cdot\bigg].
\end{align}

We now have
\begin{align}
\label{deltaexpand2}
e^{-i\theta_{\mathbf{k}+\mathbf{A}\downarrow}}\mathcal{G}^\dagger_{\mathbf{k}+\mathbf{A}}\Delta_\alpha^{\;\beta}\mathcal{G}_{\mathbf{k}-\mathbf{A}\beta}e^{i\theta_{\mathbf{k}-\mathbf{A}\uparrow}}
&=\Delta e^{i\alpha_\mathbf{k}}\bigg[1+iA^j(2\mathcal{A}_\mathbf{k}-\partial_j\theta_{\mathbf{k}\uparrow}-\partial_j\theta_{\mathbf{k}\downarrow})-2A^iA^jg_{\mathbf{k},ij}\nonumber\\
&-\frac{1}{2}A^iA^j(2\mathcal{A}_{\mathbf{k}i}-\partial_i\theta_{\mathbf{k}\uparrow}-\partial_i\theta_{\mathbf{k}\downarrow})(2\mathcal{A}_{\mathbf{k}j}-\partial_j\theta_{\mathbf{k}\uparrow}-\partial_j\theta_{\mathbf{k}\downarrow})+\frac{i}{2}A^jA^i\partial_i\partial_j\alpha_\mathbf{k}\bigg].
\end{align}

Hence, we see the familiar gauge transformation of the Berry connection: $2\mathcal{A}_\mathbf{k}-\partial_j\theta_{\mathbf{k}\uparrow}-\partial_j\theta_{\mathbf{k}\downarrow}$. The factor of two in front and the two theta terms comes from the fact that we are actually treating two Bloch functions for the two spins simultaneously. In addition, we see additional term $\frac{i}{2}A^jA^i\partial_i\partial_j\alpha_\mathbf{k}$ coming from the expansion of the exponentials \eqref{exp1} and \eqref{exp2}.

We will now show that order-by-order in powers of $\mathbf{A}$ the magnitude of the right-hand side of \eqref{deltaexpand1} is equal to the magnitude of the right-hand side of \eqref{deltaexpand2}. The magnitude squared of the right-hand side of \eqref{deltaexpand2} can be written as
\begin{align}
(\mbox{real part})^2+(\mbox{imaginary part})^2&=\bigg[1-2A^iA^j(g_{\mathbf{k}ij}-\mathcal{A}_{\mathbf{k}i}\mathcal{A}_{\mathbf{k}j})+2A^iA^j\mathcal{A}_{\mathbf{k}i}\partial_j\alpha_\mathbf{k}-\frac{1}{2}A^iA^j(\partial_i\alpha_\mathbf{k}\partial_j\alpha_\mathbf{k}-i\partial_i\partial_j\alpha_\mathbf{k})\bigg]^2\nonumber\\
&+\bigg(2A^j\mathcal{A}_{\mathbf{k}j}-A^j\partial_j\alpha_\mathbf{k}+\frac{i}{2}A^iA^j\partial_i\partial_j\alpha_\mathbf{k}\bigg)^2\nonumber\\
&=[1-2A^iA^j(g_{\mathbf{k}ij}-\mathcal{A}_{\mathbf{k}i}\mathcal{A}_{\mathbf{k}j})]^2+(2A^j\mathcal{A}_{\mathbf{k}j})^2+\mbox{terms with }\alpha_\mathbf{k}\nonumber\\
&=|\mathcal{G}^\dagger_{\mathbf{k}+\mathbf{A}}\Delta_\alpha^{\;\beta}\mathcal{G}_{\mathbf{k}-\mathbf{A}\beta}|^2+\mbox{terms with }\alpha_\mathbf{k}.
\end{align}

Let us group the terms involving $\alpha_\mathbf{k}$ in powers of $\mathbf{A}$. The zeroth and first order satisfy the self-consistent condition: $0=0$. The second order gives
\begin{align}
2A^iA^j\bigg(2\mathcal{A}_{\mathbf{k}i}\partial_j\alpha_\mathbf{k}&-\frac{1}{2}\partial_i\alpha_\mathbf{k}\partial_j\alpha_\mathbf{k}-2\mathcal{A}_{\mathbf{k}i}\partial_j\alpha_\mathbf{k}+\frac{1}{2}\partial_i\alpha_\mathbf{k}\partial_j\alpha_\mathbf{k}\bigg)=0.
\end{align}

This shows explicitly that even if we expand in powers of $\mathbf{A}$, the $\alpha_\mathbf{k}$ terms coming from the gauge transformation cancel order by order so that the magnitude of the order parameter remains invariant. Since the Higgs mode is a fluctuation of this magnitude, this shows explicitly its gauge invariance even when expanded in powers of $\mathbf{A}$.

In the pseudospin formalism, the order parameter is given by
\begin{eqnarray}
\Delta(t)=\Delta_0+\delta\Delta(t)=U\sum_\mathbf{k}(\sigma^{x(0)}_\mathbf{k}+\delta\sigma^x_\mathbf{k}+i\delta\sigma^y_\mathbf{k}).
\end{eqnarray}
In general, this is complex so that $\delta\Delta(t)$ is \textit{not} yet the Higgs mode as it contains imaginary part. The Higgs mode is the fluctuation of the magnitude with respect to the initial constant value 
\begin{eqnarray}
\label{deltah}
\Delta_H(t)=|\Delta(t)|-|\Delta_0|.
\end{eqnarray}
A gauge transformation changes the phase of $\Delta(t)$, but it is obvious from \eqref{deltah} that the Higgs mode is gauge invariant.

For simplicity, we usually perform a gauge transformation so that the zeroth order $\Delta_0$ is real. We can then write
\begin{eqnarray}
\Delta(t)=\Delta_0+\delta\Delta_R(t)+i\delta\Delta_I(t).
\end{eqnarray}

This gives the Higgs mode
\begin{eqnarray}
\Delta_H(t)=\sqrt{[\Delta_0+\delta\Delta_R(t)]^2+\delta\Delta_I(t)^2}-\Delta_0.
\end{eqnarray}

We are interested in the small fluctuations so that we can expand the square root factor and obtain
\begin{eqnarray}
\Delta_H(t)\approx \Delta_0\left(1+\frac{\delta\Delta_R(t)}{\Delta_0}\right)-\Delta_0=\delta\Delta_R(t).
\end{eqnarray}

That is, for small fluctuations $|\delta\Delta(t)|\ll\Delta_0$, the Higgs mode is simply the real part $\delta\Delta_R(t)$.


\end{widetext}

\end{document}